\documentclass{aa}

\begin{document}

\title{Magnetic reconnection and the Kelvin-Helmholtz instability in the solar corona}
\author{T. A. Howson \inst{1} \and I. De Moortel \inst{1,2} \and D. I. Pontin \inst{3}}
\institute{School of Mathematics and Statistics, University of St. Andrews, St. Andrews, Fife, KY16  9SS, U.K. \and Rosseland Centre for Solar Physics, University of Oslo, PO Box 1029  Blindern, NO-0315 Oslo, Norway \and School of Mathematical and Physical Sciences, University of Newcastle, Callaghan, NSW 2308, Australia}

\abstract{The magnetic Kelvin-Helmholtz instability (KHI) has been proposed as a means of generating magnetohydrodynamic turbulence and encouraging wave energy dissipation in the solar corona, particularly within transversely oscillating loops.}
{Our goal is to determine whether the KHI encourages magnetic reconnection in oscillating flux tubes in the solar corona. This will establish whether the instability enhances the dissipation rate of energy stored in the magnetic field.}
{We conducted a series of three-dimensional magnetohydrodynamic simulations of the KHI excited by an oscillating velocity shear. We investigated the effects of numerical resolution, field line length, and background currents on the growth rate of the KHI and on the subsequent rate of magnetic reconnection.}
{The KHI is able to trigger magnetic reconnection in all cases, with the highest rates occurring during the initial growth phase. Reconnection is found to occur preferentially along the boundaries of Kelvin-Helmholtz vortices, where the shear in the velocity and magnetic fields is greatest. The estimated rate of reconnection is found to be lowest in simulations where the KHI growth rate is reduced. For example, this is the case for shorter field lines or due to shear in the background field.} 
{In non-ideal regimes, the onset of the instability causes the local reconnection of magnetic field lines and enhances the rate of coronal wave heating. However, we found that if the equilibrium magnetic field is sheared across the Kelvin-Helmholtz mixing layer, the instability does not significantly enhance the rate of reconnection of the background field, despite the free energy associated with the non-potential field.}
\maketitle


\section{Introduction}\label{Intro}
Magnetic reconnection is a fundamental plasma process that can cause significant energy release in the solar corona \citep[see e.g.][]{Priest2000, Low2003, Pontin2012, Hesse2020}. It is likely to play a critical role in many physical phenomena in the solar atmosphere, including flares \citep[e.g.][]{Moore2001}, coronal mass ejections \citep[e.g.][]{Antiochos1999}, and small-scale heating events that maintain the high temperatures of the corona \citep[e.g.][]{Heyvaerts1984, Longcope2015}. In this context, magnetohydrodynamic (MHD) turbulence, and the associated reconnection, may be particularly important for the coronal energy budget \citep[e.g.][]{Einaudi1996, Hendrix1996, Priest1998, Cranmer2007, Vlahos2019}. 

One potential driver of turbulent flows in the solar corona is the non-linear evolution of MHD waves. In the years since the TRACE mission identified oscillations in coronal structures \citep[e.g.][]{Aschwanden1999, Nakariakov1999}, many studies have established the prevalence of wave power throughout the solar atmosphere \citep[for example, see reviews by][]{Nakariakov2005, Mathioudakis2013, Liu2014, Wang2016}. These oscillations are interesting in the context of the coronal heating problem, as they may contain sufficient energy to balance expected losses \citep[e.g.][]{McIntosh2011}. However, these measurements vary widely throughout the corona \citep[see, for example, the review by][]{VanDoors2020} and, in addition, it remains unclear whether wave energy can be dissipated on sufficiently short timescales to contribute significantly to balancing the expected coronal losses \citep{Arregui2015, Cargill2016}. 

On account of the large Reynolds numbers that are expected for typical coronal conditions, in order to attain a reasonable rate of dissipation, wave energy needs to be transferred to short length scales. Throughout the history of coronal wave research, a variety of plausible mechanisms have been proposed to achieve this. These processes include phase mixing \citep{Heyvaerts1983} and resonant absorption \citep{Ionson1978}. Alternatively, MHD turbulence may form as a result of the non-linear interaction of counter-propagating wave modes \citep[e.g.][]{Hollweg1986, VanBalle2011, Asgari-Targhi2013}. In this case, different wave modes can be associated with different drivers at the two magnetic foot points of a closed structure or by reflections of propagating waves due to density stratification. Additionally, \citet{Magyar2016b} demonstrate that the presence of transverse inhomogeneities in coronal structures can also induce the development of turbulent-like flows during oscillations.

An alternative process by which MHD waves can drive turbulent flows is via the formation of the magnetic Kelvin-Helmholtz instability (KHI) across a velocity shear layer. For example, \citet{Browning1984} presented an analytical investigation of the instability induced by phase mixing Alfv\'en waves. More recently, many authors have studied the development of this instability in numerical simulations of oscillations in coronal structures \citep[e.g.][]{Terradas2008a, Antolin2014, Magyar2015, Karampelas2017, DiazSuarez2021}. During the decay of kink oscillations, the processes of resonant absorption and phase mixing can combine to induce a large velocity shear across the boundary of magnetic flux tubes. Analytic treatments have shown that in the case of standing waves in non-twisted coronal loops, this shear is always unstable to the KHI at the wave antinodes \citep{Zaqarashvili2015}. In this regime, magnetic twist can suppress the growth rate of the instability \citep{Soler2010, Howson2017A, Terradas2018, Barbulescu2019}. In addition, the development of the instability in numerical simulations is also sensitive to spatial resolution and (user-imposed) transport coefficients \citep[e.g.][]{Howson2017B}.

Whilst detections of the KHI in the solar corona have been reported \citep[e.g.][]{Foullon2011, Ofman2011}, direct observational evidence of the KHI excited by oscillations in the boundary of coronal loops remains limited. This may be due to constraints associated with the resolving power of contemporary telescopes and is not necessarily indicative of the absence of this process. Indeed, \citet{Antolin2018}, showed that some features of observations of kink oscillations in spicules may be evidence of Kelvin-Helmholtz vortices. Additionally, the apparent heating of plasma at the boundaries of oscillating filaments \citet{Okamoto2015, Antolin2015} is potentially evidence of the mixing of hot coronal and cold prominence plasma as a result of the instability \citep{Hillier2019}. 

In terms of coronal wave heating, the energetic significance of the KHI has been scrutinised in a number of numerical and analytical studies. In simulations of continuously driven coronal loops, several authors \citep[e.g.][]{Karampelas2018, Karampelas2019, Guo2019b, Guo2019a} find turbulent flows develop throughout the loop cross section and energy is dissipated throughout the volume of the flux tube. More recently, \citet{Shi2021} presented a model that shows this heating mechanism is able to balance radiative losses for certain parameters. This recent paper presents favourable conditions for wave heating by considering a relatively low density loop (hence low radiative losses) that is excited by continuous resonant foot point driving.

Despite these positive results, it remains unclear whether wave heating models are able to dissipate sufficient energy in the corona in more general conditions. In particular, for non-resonant driving in closed loops with low energy dissipation rates, the Poynting flux injected by single frequency sinusoidal wave drivers is often relatively low \citep[e.g.][]{Howson2019, Prok2019}. Furthermore, in the context of the KHI, \citet{Hillier2020} used a simple mean-field model \citep{Hillier2019} to show that observed wave amplitudes are typically insufficient to power coronal heating through turbulent dissipation. Thus, if it is energetically relevant in the corona, the KHI must extract energy from an alternative source. A key goal of the current study is to evaluate the propensity for the KHI to dissipate free energy in the background field, instead of simply dissipating the energy in the perturbed component. In this context, free energy is used to mean the difference between the magnetic energy in the initial field and that associated with the unique, potential field with the same boundary conditions. In Sect. \ref{num_method}, we outline our model, in Sect. \ref{Sec_Res}, we present our results and, in Sect. \ref{Discussion}, we discuss the implications of our findings in the context of coronal wave research.

\section{Numerical method} \label{num_method}
For the simulations described within this article, we used the Lare3d code \citep{Arber2001}. This numerical scheme advances the full, resistive, three dimensional, MHD equations in normalised form. The equations are given by
\begin{equation}\frac{\text{D}\rho}{\text{D}t} = -\rho \vec{\nabla} \cdot \vec{v}, \end{equation}
\begin{equation} \label{eq:motion} \rho \frac{{\text{D}\vec{v}}}{{\text{D}t}} = \vec{j} \times \vec{B} - \vec{\nabla} P + \vec{F}_{\text{visc.}}, \end{equation}
\begin{equation} \label{eq:energy} \rho \frac{{\text{D}\epsilon}}{{\text{D}t}} = - P(\vec{\nabla} \cdot \vec{v}) + \eta \lvert \vec{j}\rvert^2 + Q_{\text{visc.}}, \end{equation}
\begin{equation}\label{eq:induction}\frac{\text{D}\vec{B}}{\text{D}t}=\left(\vec{B} \cdot \vec{\nabla}\right)\vec{v} - \left(\vec{\nabla} \cdot \vec{v} \right) \vec{B} - \vec{\nabla} \times \left(\eta \vec{\nabla} \times \vec{B}\right), \end{equation}
\begin{equation}\label{eq:state} P = 2 k_BnT.
\end{equation}
Here, $\rho$ is the plasma density, $\vec{v}$ is the velocity, $\vec{j}$ is the current density, $\vec{B}$ is the magnetic field, $P$ is the gas pressure and $\epsilon$ is the specific internal energy. In these equations, the resistivity, $\eta$ and viscosity, $\nu$ are included as non-ideal terms which dissipate energy from the magnetic and velocity fields, respectively. In the following results, both terms have a suppressive effect on the growth rate of the instability by reducing the cross-field velocity shear \citep{Howson2017A}. The viscosity is a sum of contributions from two small shock viscosity terms which are included within all simulations to ensure numerical stability. Together, these contribute a force, $\vec{F}_{\text{visc.}}$ on the right-hand side of the equation of motion (\ref{eq:motion}) and a heating term, $Q_{\text{visc}}$ to the energy equation (\ref{eq:energy}). The exact nature of these terms are discussed in detail in \citet{Arber2018}. Using the notation detailed in the referenced manual, we have set $\text{visc1} = 10^{-2}$ and $\text{visc2} = 5 \times 10^{-2}$.

In the following, we have initially considered simulations with zero explicit resistivity ($\eta=0$), so that any reconnection is facilitated by numerical resistivity. In Sect. \ref{non_id_reg}, we discuss the effects of an explicit anomalous resistivity which is detailed later. At this point, it is important to note that Lare3d does not force energy conservation and thus, any numerical dissipation will not contribute to plasma heating. In the simulations described here (excluding the lower resolution cases presented in Sect. 3.2), the effective (numerical) magnetic Reynolds number during the initial growth phase is approximately $10^5$. This is estimated based on the development of currents in the simulations compared with the same setup but with an explicit resistivity.

\begin{figure}[h]
  \centering
  \includegraphics[width=0.48\textwidth]{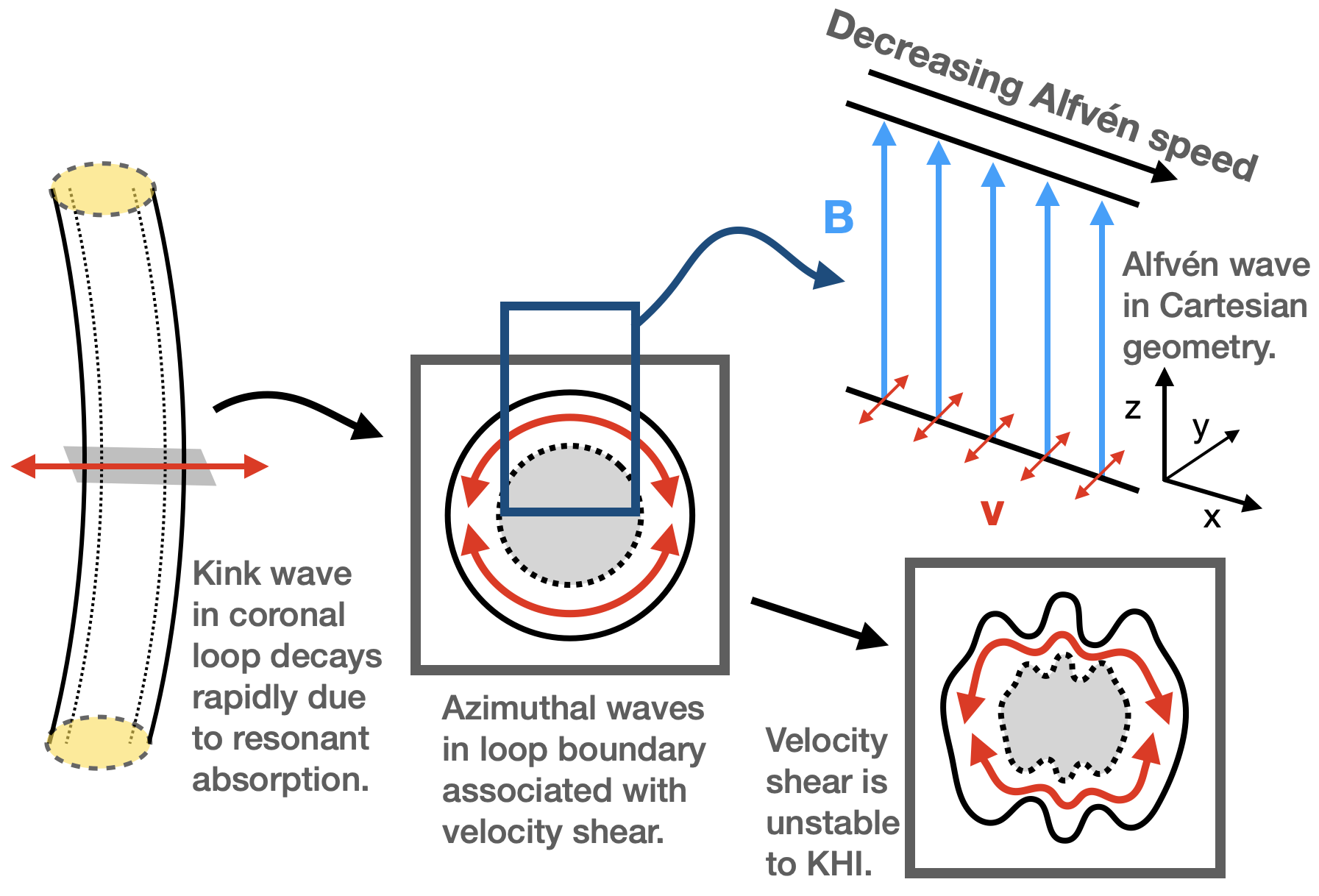}
  \caption{Schematic of the Kelvin-Helmholtz instability forming in transversely oscillating coronal loops. We also show the Cartesian geometry considered in this article.}
  \label{Cartoon}
\end{figure}

\begin{figure}[h]
  \centering
  \includegraphics[width=0.48\textwidth]{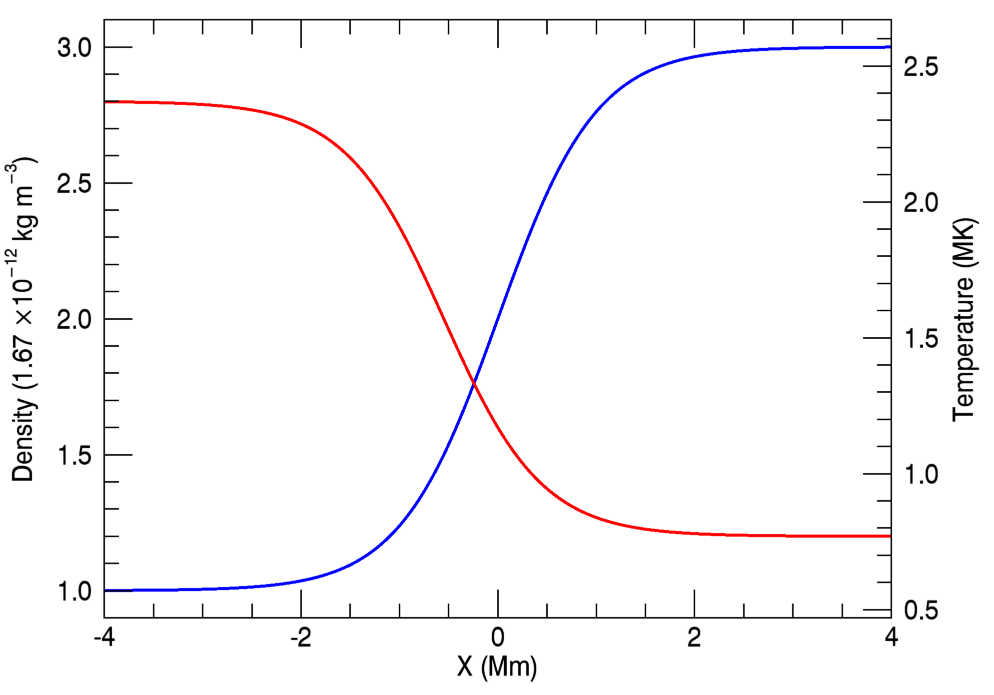}
  \caption{Transverse profiles of the initial density (blue) and initial temperature (red).}
  \label{in_dens_temp}
\end{figure}
In numerical models of kink waves in coronal flux tubes (references detailed in Sect. \ref{Intro}), an oscillating velocity shear in the boundary of the magnetic structure becomes unstable to the KHI. In this study, we model this with a very simple setup where both the axial curvature and the cylindrical form of the loop have been neglected. In Fig. \ref{Cartoon}, we show the geometry in relation to the KHI forming in the boundary of an oscillating coronal loop. With this geometry, we have simulated a plasma slab which approximates the oscillatory behaviour in the boundary of a flux tube (lower right hand panel in Fig. \ref{Cartoon}). The axial periodicity around the circumference of the flux tube is mimicked with the boundary conditions described in Sect. \ref{sect_bc}.

\subsection{Initial conditions}
We initially considered a uniform, straight magnetic field of strength 20 G (the effects of a non-potential field are described in Sect. \ref{sec:npf}), embedded in a coronal plasma. For all simulations, the numerical domain had dimensions of -4 Mm $\le x \le$ 4 Mm, -2 Mm $\le y \le$ 2 Mm and $-l/2 \le z \le l/2$. Here, $l$ is the height of the numerical domain and, hence, the length of magnetic field lines (in the potential field cases). We initially select $l = 100$ Mm and the effects of different loop lengths are considered in Sect. \ref{sec_fll}. The magnetic field was initially aligned with the $z$-axis. We imposed a density profile, $\rho$ as
\begin{equation} 
\rho(x) = \rho_0\left\{2 + \tanh\left(\frac{x}{l_0}\right)\right\}. 
\end{equation}
Here, $\rho_0 = 1.67 \times 10^{-12} \text{ kg m}^{-3}$ and $l_0 = 1$ Mm. This profile is shown in Fig. \ref{in_dens_temp} (blue curve). The temperature was defined to give $\beta = 0.04$ and to ensure $\nabla P = 0$ everywhere. The temperature had a mean value of approximately 1.4 MK. The initial temperature profile is shown in Fig. \ref{in_dens_temp} (red curve). The effects of gravity, thermal conduction and radiative losses were neglected in these simulations.

\subsection{Boundary conditions} \label{sect_bc}
\begin{figure}[h]
  \centering
  \includegraphics[width=0.48\textwidth]{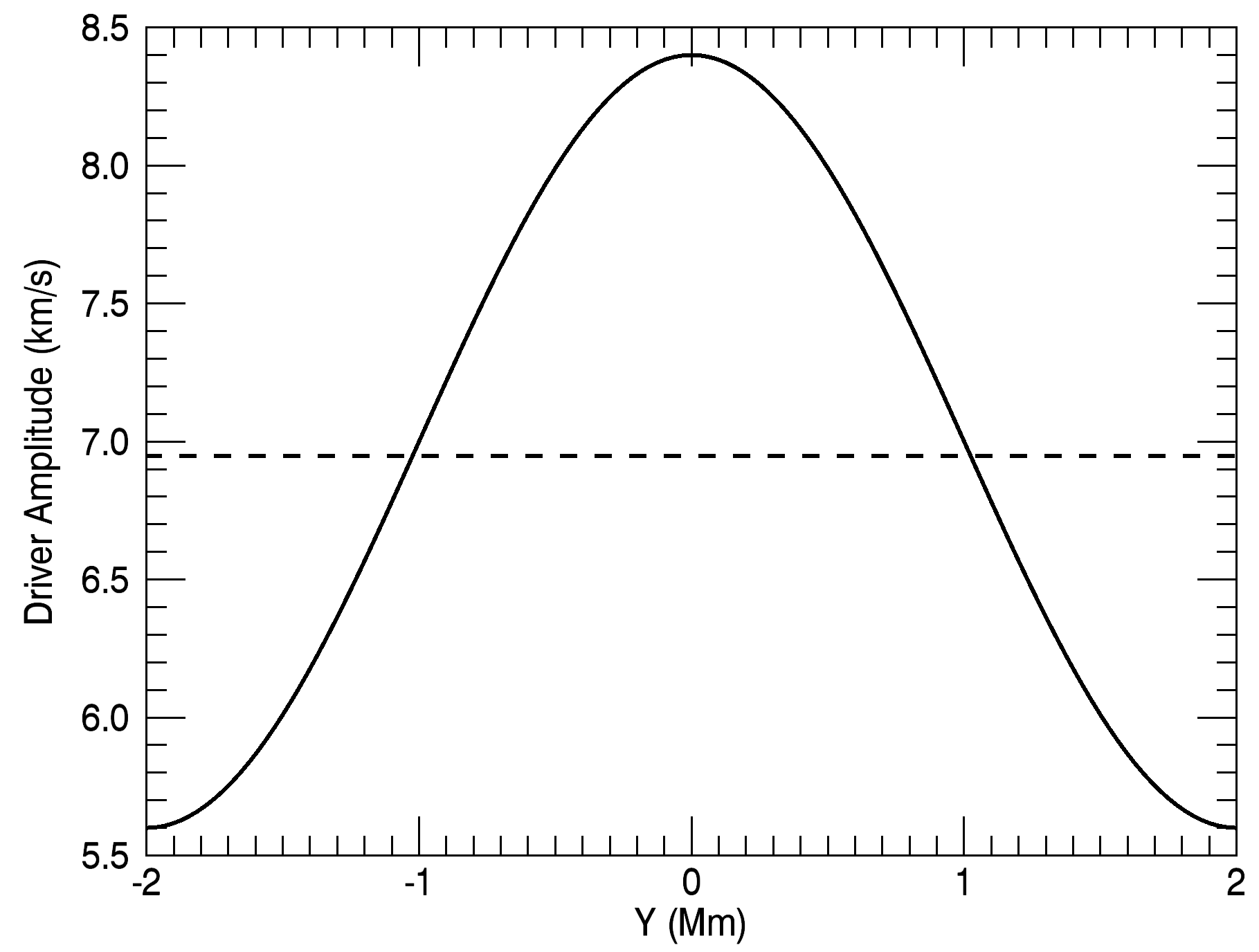}
  \caption{\emph{Solid line:} Amplitude of the driving velocity as a function of $y$. \emph{Dashed line:} Profile of $v_y(y)$ at the mid-plane of the domain when the peak of the first wave front reaches this height.}
  \label{driver_profile}
\end{figure}

In all of the following simulations, zero-gradient conditions were imposed for all variables across the $x$ boundaries and the $y$ boundaries were defined to be periodic. At the lower $z$ boundary, zero-gradient conditions were imposed for all variables except the velocity field. On this boundary, the $x$ and $z$ components of the velocity were set to 0 and transverse waves were excited by imposing a velocity profile, $\vec{v}=v_y\hat{\vec{y}}$. This wave driver was defined as
\begin{equation} \label{vy_driver}
v_y(y, t) = v_0 \sin\left(\omega t\right)\left\{1 + \Delta \cos\left(\frac{2 \pi y}{y_{\text{max}} - y_{\text{min}}}\right)\right\}.
\end{equation}
Here $v_0 \approx 7 \text{ km s}^{-1} $ is the driver amplitude and the term inside the braces is a factor which ensured that the excited waves were mildly compressible. For the following simulations, we set $\Delta = 0.2$. The variation along the $y$ axis that was induced by this term allowed the Kelvin-Helmholtz instability to form. It fulfils a similar role to random noise perturbations that are included in many numerical simulations of instabilities \citep[e.g.][]{Baty2003, Zhang2009, Mostl2013, Hillier2019}. We find that for twice the value of $\Delta$, the instability develops at the same time and the form of the vortices that form are very similar. The form of $v_y$ along the line $x=z=0$ Mm at $t=\pi/2\omega$ (time of maximum velocity) is shown in Fig. \ref{driver_profile} (solid line). The dashed line is discussed in Sect. \ref{Sec_Res}.

In equation (\ref{vy_driver}), $\omega$ is the frequency of the imposed driver and was defined as
\begin{equation} \label{freq_def}
\omega = \frac{v_{A,0} \pi}{l},
\end{equation}
where $v_{A,0}$ is the initial (uniform) Alfv\'en speed on the plane $x=0$ Mm. As a result, the frequency of the wave driver exactly matches the natural Alfv\'en frequency of all field lines contained on this plane. For $l = 100$ Mm, equation (\ref{freq_def}) gives an angular frequency of approximately 0.031 $\text{s}^{-1}$. Finally, a perfectly reflecting boundary was imposed at $z=l/2$ (upper boundary).

\section{Results}\label{Sec_Res}
We begin our analysis by considering the wave dynamics and the development of the KHI in ideal, uniform field simulations with $l = 100$ Mm. The imposed wave driver excites transverse waves that propagate upwards in the domain. As the waves propagate, they lose almost all of their compressible nature in the $y$ direction as energy is transferred across field lines in the direction of the wave polarisation. As a result, by the time the waves reach the midplane, there is almost no variation of $v_y$ along the $y$ direction (see dashed line Fig. \ref{driver_profile}). The very weak inhomogeneity that remains is sufficient to act as a seed for the Kelvin-Helmholtz instability. In addition to this effect, the cross-field gradient in the Alfv\'en speed induces phase mixing \citep{Heyvaerts1983} and encourages the formation of small scales in the velocity field. 

Upon reaching the $z=z_{\text{max}}$ boundary, the wave fronts reflect and induce a counter-propagating mode. The upward and downward propagating waves then proceed to constructively and destructively interfere with each other for the remainder of the simulations. For the field lines on $x=0$, in the linear regime, the matching of the driver frequency with the natural Alfv\'en frequencies will induce the formation of a resonance. This will lead to a large increase in the wave amplitude over the course of many wave periods, producing a standing fundamental mode on this plane. 

\begin{figure}[h]
  \centering
  \includegraphics[width=0.46\textwidth]{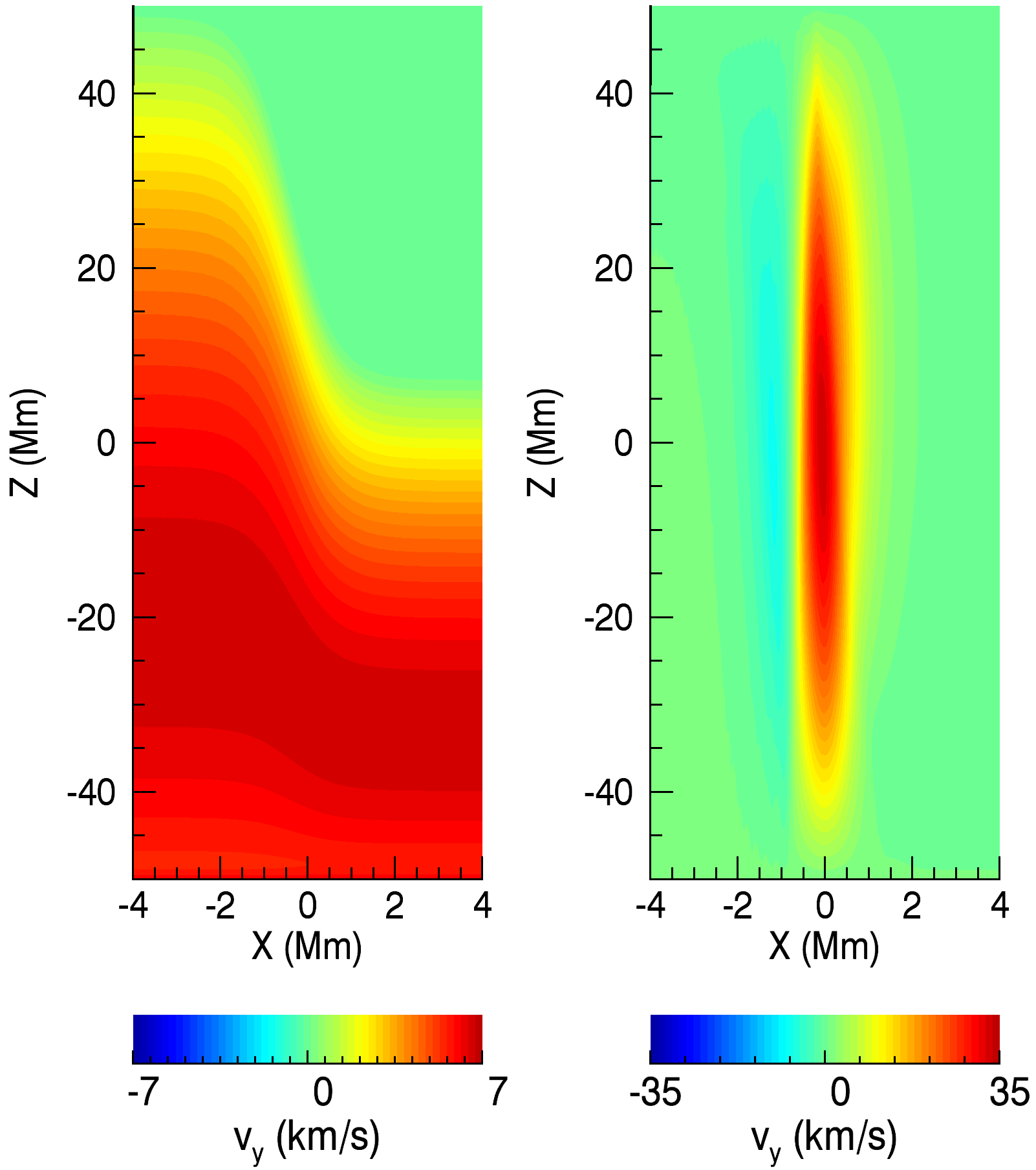}
  \caption{Wave velocity on the $y=0$ plane at $t \approx 70$ s (left) and $t \approx 500$ s (right).}
  \label{vel_profile}
\end{figure}

\begin{figure*}[h]
  \centering
  \includegraphics[width=0.9\textwidth]{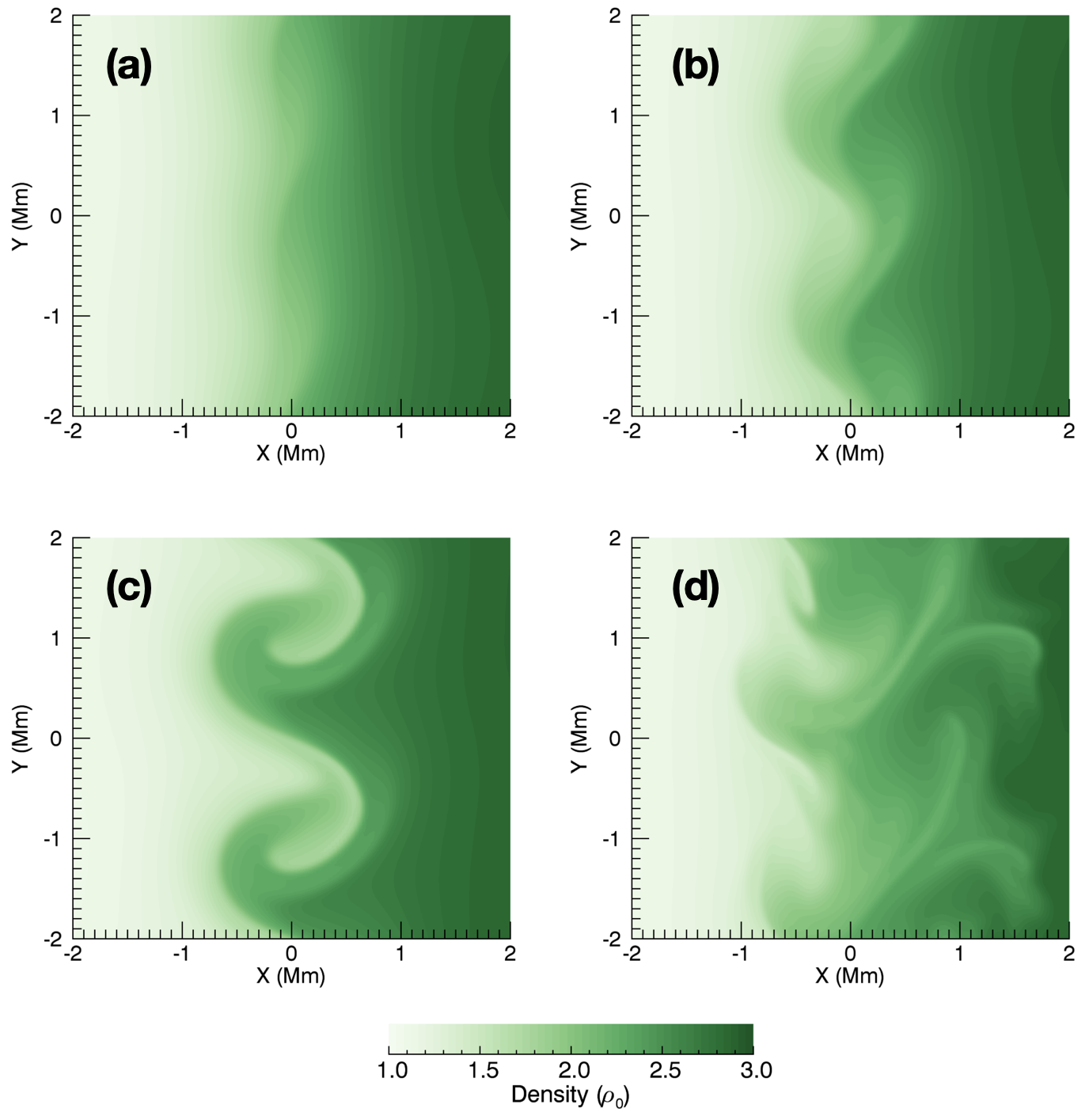}
  \caption{Evolution of the density profile at $z=0$ Mm during the development of the Kelvin-Helmholtz instability in the $l = 100$ Mm, uniform field, high resolution simulation. We show the profile at  (a) $t \approx$ 1000 s, (b) $t \approx$ 1045, (c) $t \approx$ 1090 and (d) $t \approx$ 2900 s. We have restricted the $x$ axis in each panel to $-2 \text{ Mm} \le x \le 2$ Mm.} 
  \label{khi_panels}
\end{figure*}

In Fig. \ref{vel_profile}, we show the $y$ component of the velocity in the $y=0$ plane at $t \approx$ 70 s (left) and 500 s (right). The left hand panel corresponds to a time before the wave front reaches the upper $z$ boundary and the right hand panel shows a time after several wave reflections but prior to the development of the Kelvin-Helmholtz instability (see below). In the left hand panel, we see the distortion of the propagating wave front which is characteristic of phase mixing Alfv\'en waves. Large gradients form in the perturbed velocity and magnetic fields close to $x=0$, which will (in non-ideal regimes) enhance the rate of viscous and Ohmic heating, respectively. In the right hand panel of Fig. \ref{vel_profile}, we see the growth of a high amplitude standing Alfv\'en wave on the plane of resonant field lines (note the different colour bar ranges for the two panels). Phase mixing continues across this layer and a large, oscillating velocity shear ($\partial{v_y} / \partial{x}$) forms. In low dissipation regimes, this velocity shear is unstable to the Kelvin-Helmholtz instability. 

The weak compressibility of the waves (associated with the driver profile) ensures there is a small variation along the $y$ direction of the simulation domain which acts as a seed for the formation of the KHI. After approximately 5 wave driving periods, characteristic vortices develop along the length of the resonant magnetic field lines. These are readily seen by the disruption caused to the initial density profile. In Fig. \ref{khi_panels}, we display horizontal cuts of the density profile in the $z=0$ plane at four stages during the development of the instability. In less than the duration of a wave period (which is approximately 200 s), small perturbations in the transverse density profile (a) and (b) develop into the large vortices (c) that are characteristic of both the hydrodynamic and magnetised forms of the instability.

Ultimately, these large vortices break down and a turbulent-like regime develops (panel (d) of Fig.  \ref{khi_panels}). By this stage, small scales are also present in the velocity and magnetic fields and as a result, the KHI will  enhance the rate of wave energy dissipation across the velocity shear layer. Since the amplitude of the standing Alfv\'en waves (in terms of the velocity field) is largest at the apex of field lines ($z=0$), the Kelvin-Helmholtz vortices are largest here. However, the density vortices form along much of the length of the field lines, with the exception of the wave nodes at the upper and lower $z$ boundaries. However, even at these locations, evidence of the instability can be seen in the perturbed magnetic field. Since the small scales in the velocity field predominantly form at the antinode, viscous heating will dominate in the $z=0$ layer. Conversely, gradients in the magnetic field are largest at the field line foot points and thus Ohmic heating will dominate closer to both the driven and reflective $z$ boundaries \citep[e.g.][]{VanDoors2007, Karampelas2017}.

\subsection{Poynting flux} \label{Sect_pf}
In order for MHD wave models to be interesting in the context of the coronal heating problem, imposed wave drivers must inject sufficient energy to balance a significant fraction of losses due to thermal conduction and optically thin radiation, for example. In ideal regimes which allow for the reflection of wave modes at a closed upper boundary, a wave driver will not continuously inject energy unless a field line is driven resonantly. Indeed, for non-resonant field lines, a wave driver is as likely to remove energy from a reflected wave front as it is to introduce new energy to the domain. As such, significant wave energy can only be introduced into a simulation domain if field lines are driven resonantly (on the $x=0$ plane in this case) or there is sufficient dissipation such that the amplitude of a reflected wave is noticeably smaller than the amplitude of the imposed driver. In this paradigm, two aspects of the KHI are particularly interesting. Firstly, as the density profile is deformed, the resonant layer and the energy injection rate are modified, and secondly, in non-ideal cases, the instability will enhance the rate of wave energy dissipation.

In these simulations, the only energy injected into (or removed from) the domain is associated with the Poynting flux through the lower $z$ boundary. Prior to any reflected waves reaching the lower boundary, at a particular point, the time averaged Poynting flux, $S(y)$, can be expressed as
\begin{equation} \label{early_PF}
S(y) = \left \langle\frac{\vec{E} \times \vec{B} \cdot \vec{\hat{z}}}{\mu_0} \right \rangle_{t}= v_A \rho v_{0}^2 (y), 
\end{equation}
where $v_A = v_A(x)$ is the local Alfv\'en speed, $\rho= \rho(x)$ is the local density and $v_{0}(y)$ is the maximum driver amplitude at a particular value of $y$ (see equation \ref{vy_driver}). We now define $\tau$ to be the period of the wave driver (also the period of the fundamental standing mode on the $x=0$ plane). As the largest value of the background Alfv\'en speed is $\sqrt{2}$ larger than the speed on the $x=0$ plane, the first reflected wave reaches the driven boundary at $t = \tau/\sqrt{2}$. Therefore, for $t > \tau/\sqrt{2}$, reflected waves will be interacting with the imposed driver and equation (\ref{early_PF}) is no longer valid. For the resonant field lines at $x=0$, the instantaneous Poynting flux will grow linearly in time until a non-linear regime is reached and the natural frequency of the field lines detunes from the driver frequency. This is often associated with the ponderomotive force redistributing density along field lines, however, in these simulations it is typically associated with the change of field line length and modification of the density profile that are associated with the formation of the KHI.

\begin{figure}[h]
  \centering
  \includegraphics[width=0.48\textwidth]{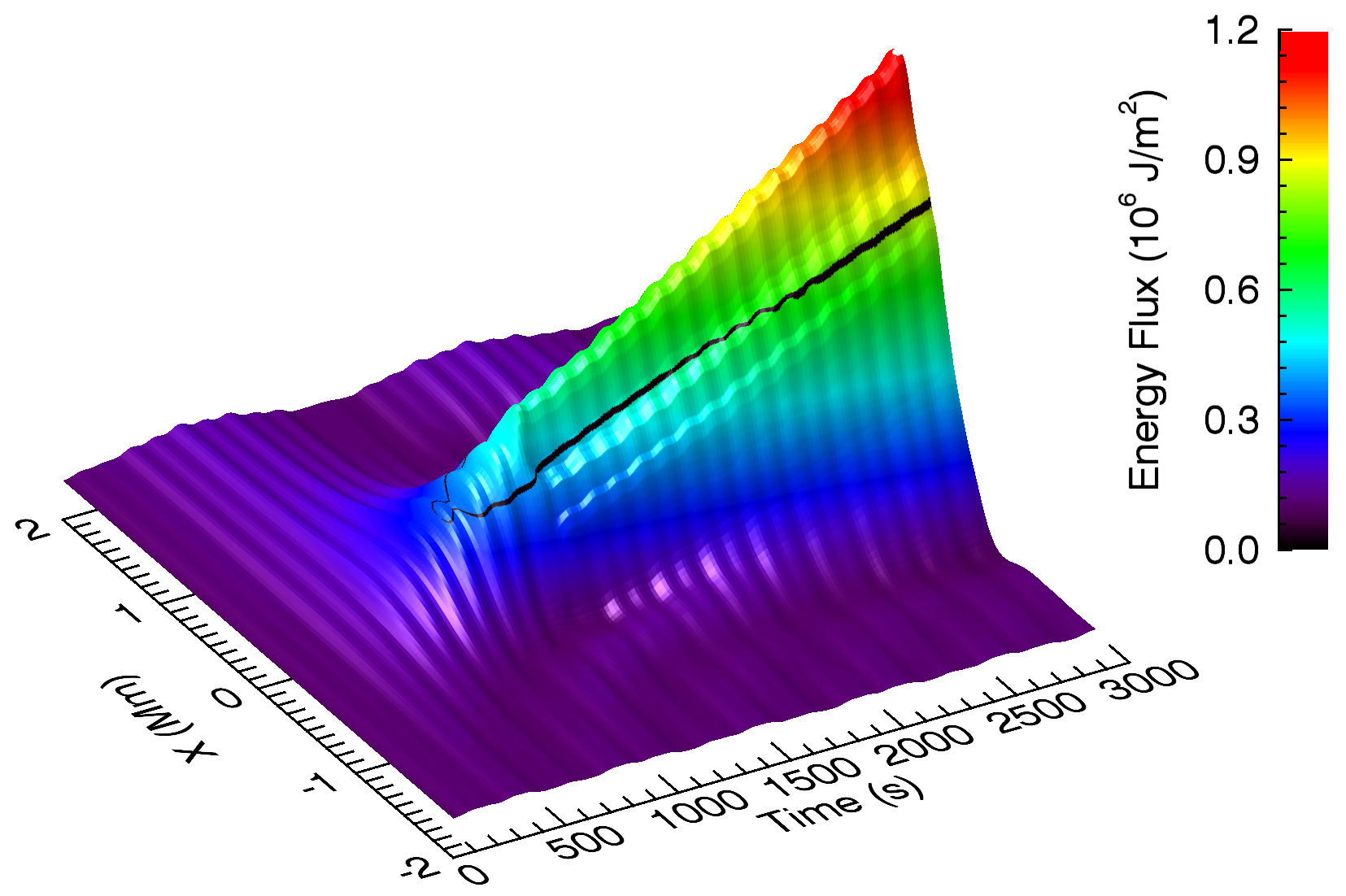}
  \caption{ Cumulative Poynting flux injected through the lower $z$ boundary by the imposed wave driver. The black contour contains points which receive sufficient energy to balance expected losses in a typical quiet region of the solar corona.}
  \label{cumulative_energy_flux}
\end{figure}

\begin{figure}[h]
  \centering
  \includegraphics[width=0.48\textwidth]{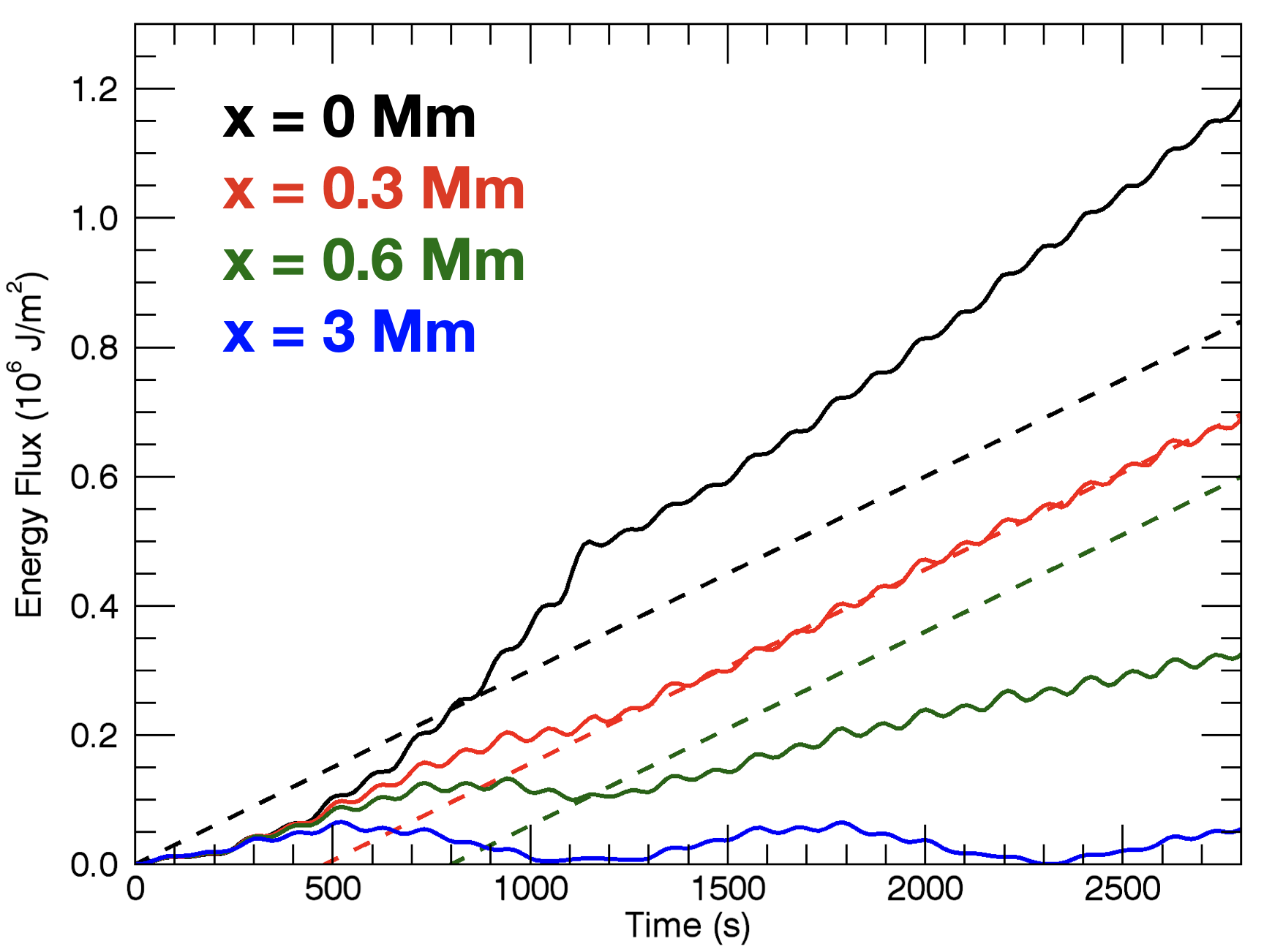}
  \caption{\emph{Solid}: Cumulative Poynting flux injected through the lower $z$ boundary by the imposed wave driver averaged along $y$ for various values of $x$. \emph{Dashed}: Estimated energy requirements in typical Quiet Sun conditions. The losses shown by the red dashed line are delayed to show that the mean Poynting flux for $x=0.3$ Mm is sufficient to balance expected Quiet Sun losses once the Kelvin-Helmholtz instability forms. Conversely, the green dashed line shows that the mean Poynting flux for $x=0.6$ Mm is not sufficient even once the instability has developed.}
  \label{cumulative_energy_flux_line}
\end{figure}

In Fig. \ref{cumulative_energy_flux}, we show the cumulative Poynting flux injected into the domain as the simulation progresses. As the invariance in the $y$ direction is destroyed during the development of the KHI, we average out the effects of the Kelvin-Helmholtz vortices by integrating along the $y$ axis. This gives the Poynting flux as a function of $x$ and $t$. The black contour contains the points that receive sufficient energy from the wave driver to balance expected losses in the Quiet Sun according to the estimates detailed in \citet{Withbroe1977}. We note that the Poynting flux injected into the vast majority of the domain is not sufficient to balance even this relatively low loss rate (compared to requirements for typical active regions).

For the purpose of clarity, in Fig. \ref{cumulative_energy_flux_line}, we show the cumulative Poynting flux through the lower $z$ boundary for different values of $x$. The solid black line ($x=0$ Mm) corresponds to the foot points of the resonant field lines and coincides with the largest average Poynting flux (see Fig. \ref{cumulative_energy_flux}). We also include the expected energy requirements for the Quiet Sun (dashed black line). Prior to the formation of the KHI at $t \approx 1000$ s, we see that the total energy is increasing approximately quadratically as the instantaneous Poynting flux (time derivative of the solid line in Fig. \ref{cumulative_energy_flux_line}) is increasing linearly (due to the relatively low wave amplitudes). Without the increase in Poynting flux due to the formation of an MHD wave resonance, the injected energy would be below Quiet Sun requirements everywhere in the domain (for $t \lesssim450$ s the gradient of the solid line is less than the gradient of the dashed line). As such, for these simulations, very efficient wave dissipation mechanisms (such that the energy release time scale was shorter than a wave period) would not be able to balance coronal losses as they would prevent a resonance developing. Again, we note that in all locations (even on resonant field lines), the energy injection rate is much lower than that required to balance expected losses in active regions \citep{Withbroe1977}.

Once the velocity shear across the resonant layer triggers the onset of the KHI, the instantaneous Poynting flux decreases. In the linear regime, and prior to the formation of the KHI, the field lines contained in the $x=0$ Mm plane are perfectly resonant. However, any change in field line length or in the local Alfv\'en speed will detune the natural frequency of the field lines from the frequency of the wave driver. When the KHI forms, many field lines on this resonant layer will increase in length and the local density (and hence Alfv\'en speed) will be modified. As such, the KHI reduces the rate of energy injection on this layer. Despite this, significant energy injection (comparable to Quiet Sun requirements) continues on this layer as the change in field line length and mean density is relatively small. We note that the decrease does not occur at the exact formation time of the Kelvin-Helmholtz vortices as it takes at least a quarter of a wave period for the reflected waves to contain information about the vortices forming at the apex of the magnetic field lines. Indeed, from this point forward the short evolution timescales of the turbulent-like flows (in comparison to the length of a wave period) will ensure that the nature of the resonant layer cannot be predicted using an instantaneous snapshot of the magnetic field and density profiles.

In the linear regime, the rate of energy injection onto non-resonant field lines will oscillate with a low frequency that reflects the difference between the driver frequency and their natural frequency. This can be seen in the blue line of Fig. \ref{cumulative_energy_flux_line} which shows the cumulative energy flux for the $x=3$ Mm plane. We also see the start of this beating behaviour on the $x=0.3$ Mm (solid red) and $x=0.6$ Mm (solid green planes). However, following the onset and growth of the KHI, there are sufficiently many (possibly temporarily) resonant field lines to substantially enhance the energy injected by the driver. In the case of $x=0.3$ Mm, we see that following the onset of the instability, the increase in energy injection is sufficient to balance Quiet Sun losses (compare solid and dashed red lines). For the weaker distortion apparent at $x=0.6$ Mm, on the other hand, the increase in the averaged Poynting flux would not be sufficient to be the only source of coronal heating (compare solid and dashed green lines). Even with the large disruption of the resonant layer, most of the computational domain ($|x| \gtrsim 0.3$ Mm) still experiences low energy injection rates. 

\subsection{Numerical resolution} \label{num_res_study}
It has been well established that dissipative effects within numerical simulations inhibit the growth rate of the Kelvin-Helmholtz instability \citep[e.g.][]{Howson2017B}. This is the case for both user-imposed transport coefficients and for numerical dissipation associated with the use of a finite difference scheme. As such, the growth rate of the instability may be artificially inhibited in many large scale MHD simulations of the solar corona. In this section, we introduce the metrics used within this article whilst examining the effects of numerical resolution in this Cartesian geometry.

\begin{figure*}[h]
  \centering
  \includegraphics[width=0.95\textwidth]{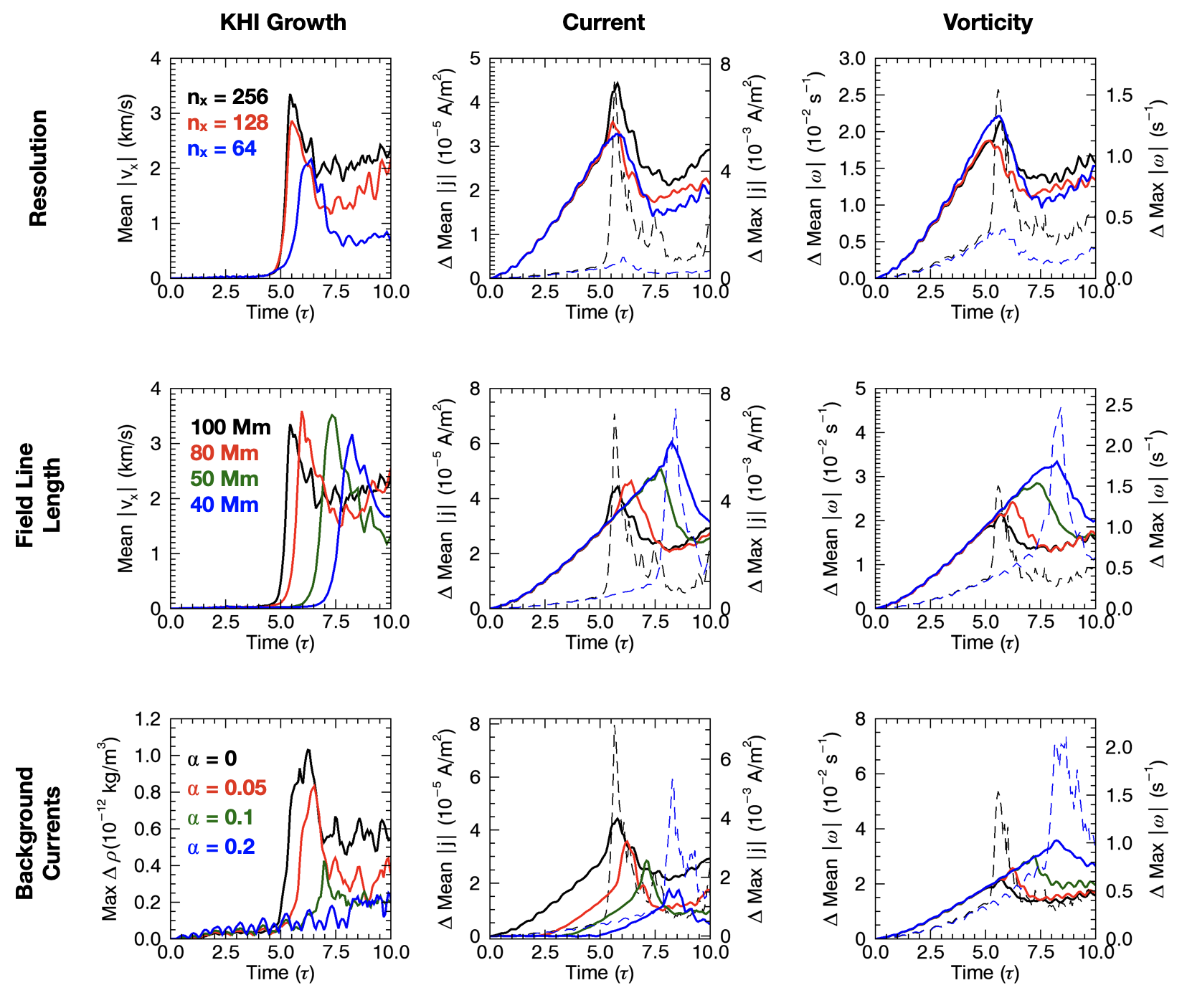}
  \caption{Growth rates of the KHI and small-scale formation for the numerical simulations considered within this article. We show the effects of numerical resolution, field line length and the inclusion of a shear component in the background field for the first, second and third rows respectively. The first column shows the growth rate of the instability (details in text). The second and third column show the change in the mean (solid lines, left hand axis) and the change in the maximum  (dashed lines, right hand axis) of the current density and the vorticity, respectively. In all cases, the change is calculated relative to the initial conditions.}
  \label{growth_panels}
\end{figure*}

\begin{figure*}[h]
  \centering
  \includegraphics[width=0.95\textwidth]{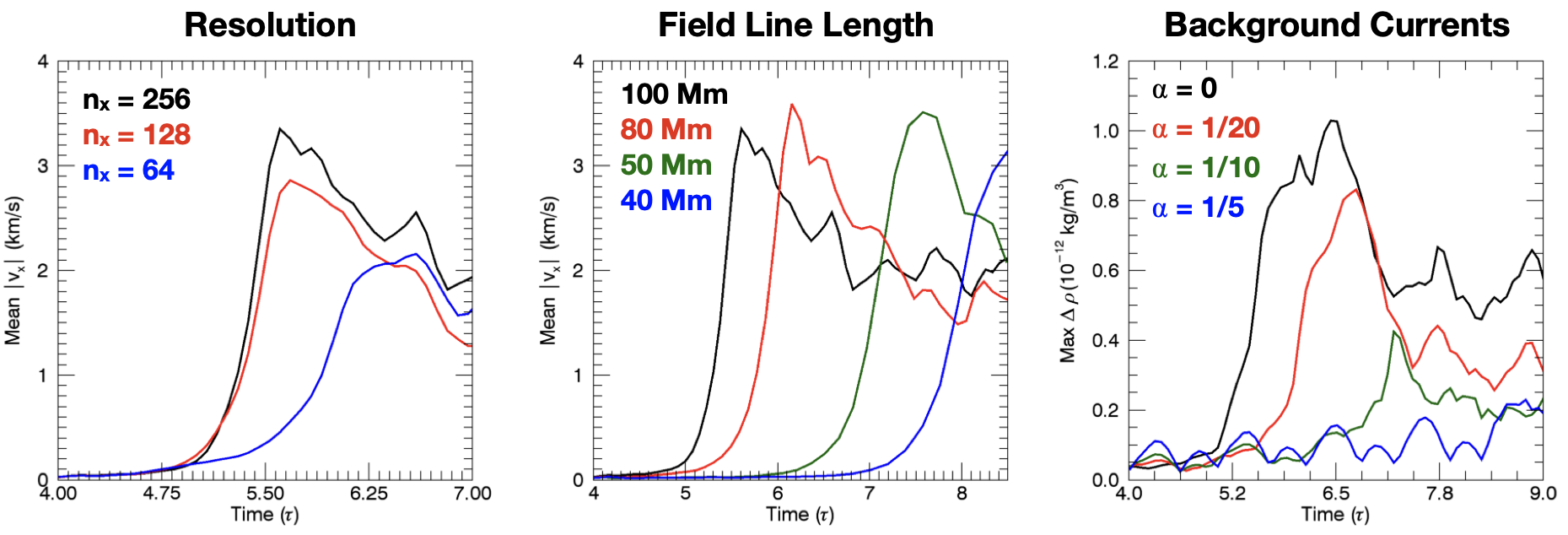}
  \caption{Zoomed in versions of the left hand column of Fig. \ref{growth_panels} showing the growth phase of the instability for different simulations.}
  \label{growth_panels_zoom}
\end{figure*}

We consider three simulations with 64, 128 and 256 grid points in the $x$ direction ($n_x$) and 32, 64 and 128 grid points in the $y$ direction ($n_y$), respectively. We note that the total length of the $x$ axis is twice the length of the $y$ axis, so, in each case the $x$ and $y$ resolutions are identical. The $z$ resolution is unchanged in each case. The highest resolution corresponds to the simulation described in Sect. \ref{Sect_pf}.

In order to evaluate the onset time and initial growth rate of the instability in each case, we note that the initial disturbance in the density profile is associated with a growth in $|v_x|$ at the antinode of the standing wave. Prior to the onset of the instability, this component of the velocity is extremely small as the driven wave is polarised in the $y$ direction. However, this quantity increases dramatically as the first Kelvin-Helmholtz vortices form. In the top left hand panel of Fig. \ref{growth_panels}, we plot the time evolution of the mean of this quantity averaged over the numerical domain for the three different resolutions. Note, for all panels in this figure, the time axis is shown as the number of periods of the wave driver ($\tau$). In each case, this quantity clearly identifies the onset time of the instability at approximately $5 \tau$. The sharp increase in the mean of $|v_x|$ corresponds to the linear growth phase of the instability and, for clarity, we show a zoomed in version of this growth phase, in the left hand panel of Fig. \ref{growth_panels_zoom}. However, this growth is not sustained and breaks down as the initial vortices collapse (see Fig. \ref{khi_panels}). As the KH-vortices fold back on themselves, the magnitude of the $x$ component of the velocity decreases. Subsequently, as a turbulent-like regime develops, later vortices are much smaller and thus velocities do not attain the same maxima again. We note that both the formation time and subsequent growth rate of the instability will be suppressed by low numerical resolution (e.g. compare black and blue curves).

The KHI is interesting in the context of coronal heating as the formation of small scales in the magnetic and velocity fields will enhance the rate of wave energy dissipation. We measure the small scales in each of these fields using the current density, $\vec{j}$, and vorticity, $\vec{\omega}$, respectively.  In the central and right-hand columns of Fig. \ref{growth_panels}, we show the evolution of these quantities for each of the resolutions considered. As the initial currents are non-zero in some of the following simulations, we consider the change in the currents and vorticities, relative to the initial conditions. In each case, solid lines represent the change in the mean value and the dashed lines show the change in the maximum value obtained within the domain. In each plot, the left hand axes correspond to the means and the right hand axes correspond to the maxima. For clarity, we only show the two extreme cases (e.g. for the top row $n_x = 64, 256$) for the maxima curves. We consider the second and third rows of Fig. \ref{growth_panels} in subsequent sections.

For all variables, we see an increase prior to the formation of the instability. This corresponds to the progression of classical phase mixing \citep{Heyvaerts1983} due to the cross-field gradient in the local Alfv\'en speed. The small scales that form across the high amplitude wave at the resonant layer are particularly important and they provide the dominant contribution in these curves. We note that before the KHI onset time, both the current and vorticity plots are similar across resolutions, suggesting that even in the low resolution $n_x = 64$ case, the early wave dynamics are well-resolved. 

The onset of the instability is associated with a large increase in the maximum currents and vorticities, particularly in the high resolution simulation (dashed black lines). These small scales form along the largest vortices where the greatest disruption to the velocity field occurs. In non-ideal regimes, these small spatial scales will be associated with high, but very localised and temporally intermittent, energy dissipation. We see that the sizes of the largest currents and vorticities are sensitive to the numerical resolution. This is not surprising as finer grids allow larger gradients to form. However, we also see greater growth rates in high resolution simulations. Following the initial rise of the maximum current and vorticity, the development of a turbulent-like regime (see panel (d) in Fig. \ref{khi_panels}), prohibits the continued existence of such large gradients. Instead, the smaller vortices are less energetic and small scales are more widely distributed throughout the mixing plasma.

Despite the rapid increase in the largest currents and vorticities during the early stages of the KHI, the mean values of these variables show little or no increase during this time. In this regime, where the KHI only develops in a relatively small proportion of the domain, we do not see a large increase in heating over most of the simulation volume. Indeed, the instability actually reduces these measures of small scales for $t \gtrsim 6 \tau$. Prior to the formation of the instability, for all resolutions, the mean current density and the mean vorticity are growing approximately linearly. This is associated with the growth of the wave amplitudes observed at the resonant layer. In non-ideal regimes, for $t < 5 \tau$, the location of energy dissipation will predominantly be across the narrow sheet of large gradients at $x=0$ Mm. Here, the instability disrupts this resonant layer and the Kelvin-Helmholtz vortices act to reduce the coherent gradients on this plane.
 
\subsection{Field line length} \label{sec_fll}
In this section, we consider the effects of field line length on the development of the magnetic KHI. We consider simulations with $l \in \left\{20, 40, 50, 80, 100\right\}$ Mm. The $l = 100$ Mm is the first simulation discussed in Sect. \ref{Sec_Res}. In general, for finite field line lengths, magnetic tension will act to suppress the growth rate of the instability as the Kelvin-Helmholtz vortices distort the magnetic field. For shorter field line lengths, the same apex displacement of a field line will produce a larger tension force. As such, we should expect the KHI to be somewhat suppressed in shorter coronal loops. In order to understand the instability for different field line lengths in our numerical simulations, we begin by considering the factors that directly affect the onset and development of the KHI (for example, the magnitude of the velocity shear and the suppressive effects of magnetic tension). This discussion will allow the instability and small-scale growth rates to be analysed in more detail.

Many previous analytical studies \citep[e.g.][]{Roberts1973, Browning1984, Soler2010, Hillier2019a, Barbulescu2019} have identified instability criteria for non-steady flows in coronal plasmas. With this in mind, we can compare our relatively complex setup, with a similar model that permits an analytic treatment and considers density and oscillatory velocity profiles which are uniform (but different) on each side of a discontinuity. The profiles are given by
\begin{equation}
\rho(x) = \begin{cases}
\rho_{-} & \text{if}\ x \le 0, \\
\rho_{+} & \text{if}\ x > 0,
\end{cases}
\end{equation}
and a velocity profile $\vec{v} = (0, v_y, 0)$ where
\begin{equation}
v_y(x) = \begin{cases}
v_{-}\cos{\omega t} & \text{if}\ x \le 0, \\
v_{+} \cos{ \omega t} & \text{if}\ x  > 0.
\end{cases}
\end{equation}
Here, $\rho_{+,-}$ and $v_{+,-}$ are constants. This setup is unstable to the KHI if
\begin{equation} \label{instability_criteria}
(\Delta v)^2 > \frac{4 k_z^2  \left(\rho_{+} + \rho_{-}\right) v^2_{A,+}}{\rho_{-}k_y^2},
\end{equation}
where $\Delta v = v_{+} - v_{-}$, $v_{A,+}$ is the background Alfv\'en speed for $x > 0$ and $k_{y,z}$ are the wave numbers of an assumed perturbation in the $y$ (parallel to wave polarisation) and $z$ (parallel to background magnetic field) directions, respectively \citep[adapted from equation 39 in][]{Hillier2019a}. 

Whilst this simplified set-up does not apply directly to the continuous velocity shear considered in this article, it is able to highlight some key criteria which are relevant for the growth rates discussed below. In particular, the $k_z$ term is affected by the field line length. As a simple comparison with our model, we can take $\rho_{+} = 3\rho_{-}$, $k_y^{-1} = y_{\text{max}} - y_{\text{min}}$ (see driver in eq. \ref{vy_driver}) and $k_z^{-1} = 2l$ as the wavelength is twice the height of the domain for a fundamental standing mode. We can also define $v_{A,+}$ using the Alfv\'en speed at $x=x_{\text{max}}$. Under these assumptions, we immediately see that the instability criterion is easily reached for $k_z \ll k_y$ (i.e. for long field lines) and further, any velocity shear is unstable for infinitely long field lines ($k_z \to 0$).

In this simple comparison with our model, the correct form for $\Delta v$ remains unclear and there are two important factors to consider. Firstly, the continuous resonant driving of the $x=0$ plane enhances the amplitude of the central field lines in comparison to the neighbouring field lines. As a result, the velocity shear ($\sim \Delta v$) increases as the simulation progresses. Secondly, the continued phase mixing across the inhomogeneous layer that occurs during the simulations creates large gradients in the velocity field. This also effectively enhances the magnitude of the shear ($\sim \Delta v$). The effects of this second factor are considered by \citet{Browning1984}.

Despite this complexity, we can constrain the effects of field line length on $\Delta v$ as follows. The apex velocity of a resonantly driven, linear, fundamental, standing Alfv\'en wave is given by \citep[e.g.][]{Prok2021}
\begin{equation} \label{v_an_amp}
v(t) = v_d \left(m' + 1\right) \sin\left\{\omega\left(t - \frac{l}{2v_{A,0}}\right)\right\}, 
\end{equation}
where $v_d$ is the amplitude of the imposed driver ($v_0$ in this case) and
\begin{equation}
m' = \left \lfloor \frac{tv_{A,0}}{l} - \frac{1}{2}\right \rfloor.
\end{equation}

\begin{figure}[h]
  \centering
  \includegraphics[width=0.45\textwidth]{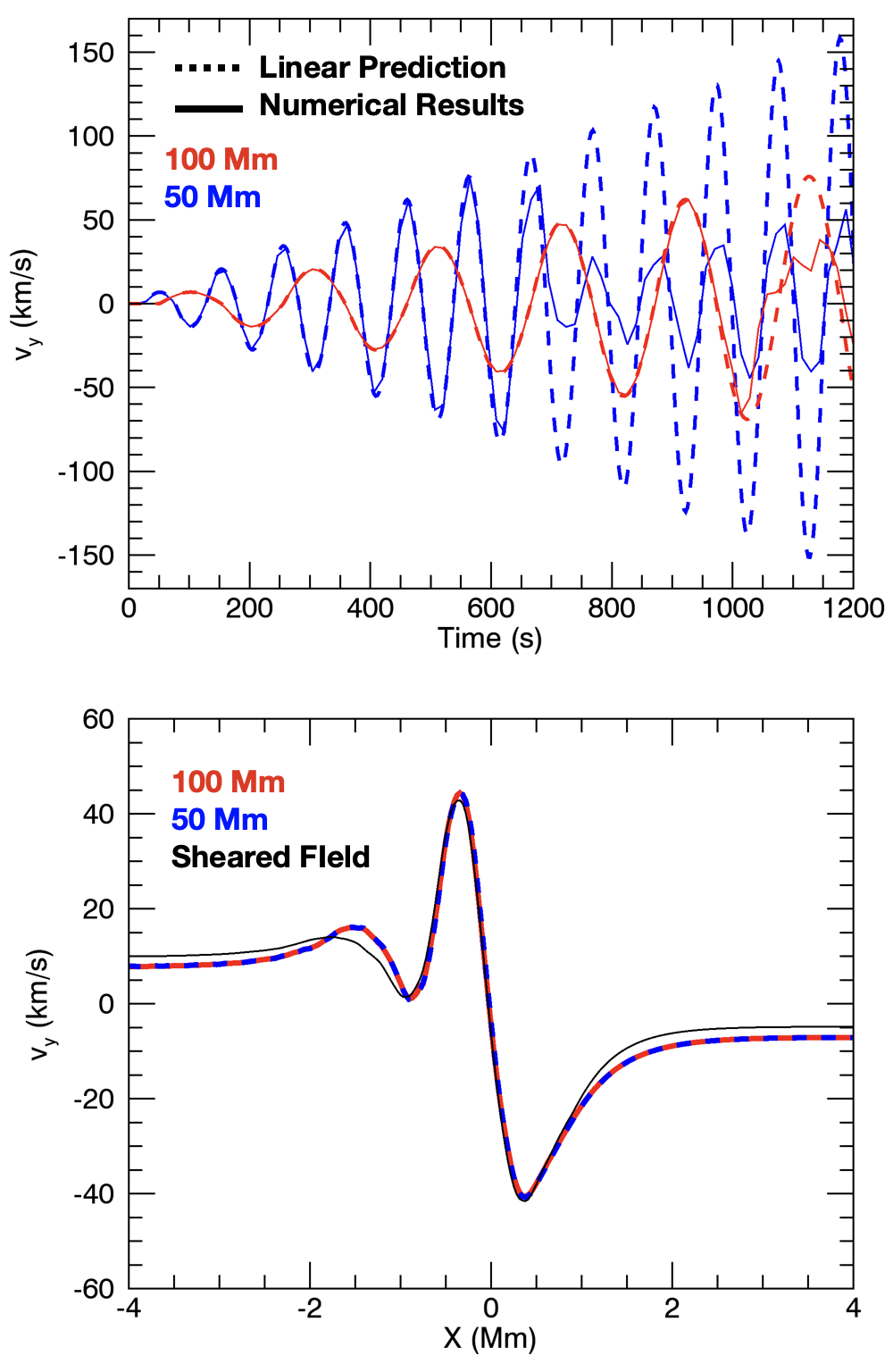}
  \caption{\emph{Upper}: Amplitudes of resonant wave for $l = 50$ Mm (blue) and $l=100$ Mm (red). Dashed lines show prediction from eq. (\ref{v_an_amp}) and solid lines are results from numerical simulations. \emph{Lower}: Cross-field phase mixing profile for $l = 50$ Mm at $t = 475$ s (blue) and $l=100$ Mm (red) at $t = 950$ s (red). The thin black line corresponds to a simulation with shear in the background field and is discussed in Sect. \ref{sec:npf}.}
    \label{something}
\end{figure}

Note, $m' + 1$ is the number of times the leading wave front has passed the $z=0$ plane (including after reflections). In the upper panel of Fig. \ref{something}, we show the amplitude of the standing wave observed at $x=y=z=0$ Mm as a function of time for the $l= 50$ Mm (blue) and $l=100$ Mm (red) cases. The solid lines show the velocities obtained in the numerical simulations and the dashed lines show the predictions given by eq. \ref{v_an_amp}. Prior to the formation of the KHI at $t\approx650$ s (blue) and $t\approx 1000$ s (red), the analytic prediction and numerical results match well. We note that there are small deviations at times of maxima velocity which are associated with a lack of temporal resolution in the output of the MHD simulations (not the simulation time step) and weak viscous effects (both numerical and due to the user-imposed shock viscosities). Following the onset of the instability, however, the simulation velocity is much smaller than the analytic formula because the resonance is disrupted.

Although the increase in wave amplitude is faster for the shorter loop, if we normalise for the Alfv\'en travel time in each case, the resonance growth occurs at the same rate. In other words, after the same number of driving periods ($\tau = 2l/v_{A,0}$), the wave amplitude is the same in both cases (prior to KHI onset). Therefore, we do not expect this effect to modify the instability onset time in terms of the number of wave periods (although it will when measured in seconds) for field lines of different lengths. 

In addition to the magnitude of the velocity difference across the resonant layer, the transverse scales in the velocity profile are also critical for the growth rate of the instability. As we have discussed, these are continuously modified by the progression of phase mixing in the simulations. However, as the cross-field gradient in the local Alfv\'en speed is unchanged between simulations, if we again normalise for the driving period, we expect phase mixing to be unchanged across simulations (prior to KHI onset). In the lower panel of Fig. \ref{something}, we show the profile of $v_y$ as a function of $x$ after approximately 4 wave periods for the $l = 50$ Mm (red) and $l = 100$ Mm (dashed blue) simulations. Indeed, we see that the velocity profile, and thus, the cross-field velocity shear are almost identical. The black line is discussed in Sect. \ref{sec:npf}.

Therefore, as a function of $\tau$, the velocity amplitude and cross field shear are independent of field length prior to the development of the KHI. As a result, any suitable proxy for $\Delta v$ in (\ref{instability_criteria}) will not change between these simulations. Therefore, due to the $k_z^2 = 1/4l^2$ term in (\ref{instability_criteria}), we expect the short field line simulations to require a greater number of wave periods to reach the threshold for instability. In the left hand panel of the middle row in Fig. \ref{growth_panels}, we see that this is indeed the case. Using the same proxy for onset time and initial growth rate as for the numerical resolution study (see Sect. \ref{num_res_study}), we see that the KHI onset is delayed by a number of wave periods for the shorter field line cases. However, it is important to note, that due to the shorter wave period, the KHI still occurs sooner in shorter field line cases (when onset time is measured in seconds). Again, for clarity, we show a zoomed in version of the linear growth phase for the different cases, in the central panel of Fig. \ref{growth_panels_zoom}. As a result of the different instability onset times, we expect any energy release associated with the instability to occur earlier for shorter field line simulations in non-ideal regimes.

As with the numerical resolution parameter space, in the central and right-hand panel of the second row in Fig. \ref{growth_panels}, we show the time evolution of the current density and vorticity for different values of $l$. In all cases, prior to the onset of the instability, there is very little difference in these quantities between simulations. Even though field aligned gradients are larger for the shorter field line cases, the cross-field, phase mixing gradients are so dominant that this has negligible effects on the magnitude of $\vec{j}$ and $\vec{\omega}$. 

As the KHI leads to a decrease in the volume integrated currents and vorticities (potentially following a small initial increase - as discussed above), the increased number of wave periods before the instability is triggered for small values of $l$, means that short field line simulations will sustain smaller scales in both $\vec{B}$ and $\vec{v}$ for a time. As long as the KHI is delayed, the continued existence of the resonance layer and the progression of phase mixing will ensure that gradients increase further across the $x=0$ Mm plane. Only once the instability induces a turbulent-like regime in all cases do the currents and vorticities return to similar values across simulations. As the maximum currents and vorticities are larger for small values of $l$, in a non-ideal regime we expect an enhancement in the maximum heating rate for shorter field lines. This is simply an effect of the enhanced time (in terms of wave periods) for which a coherent resonance exists.

\begin{figure}[h]
  \centering
  \includegraphics[width=0.48\textwidth]{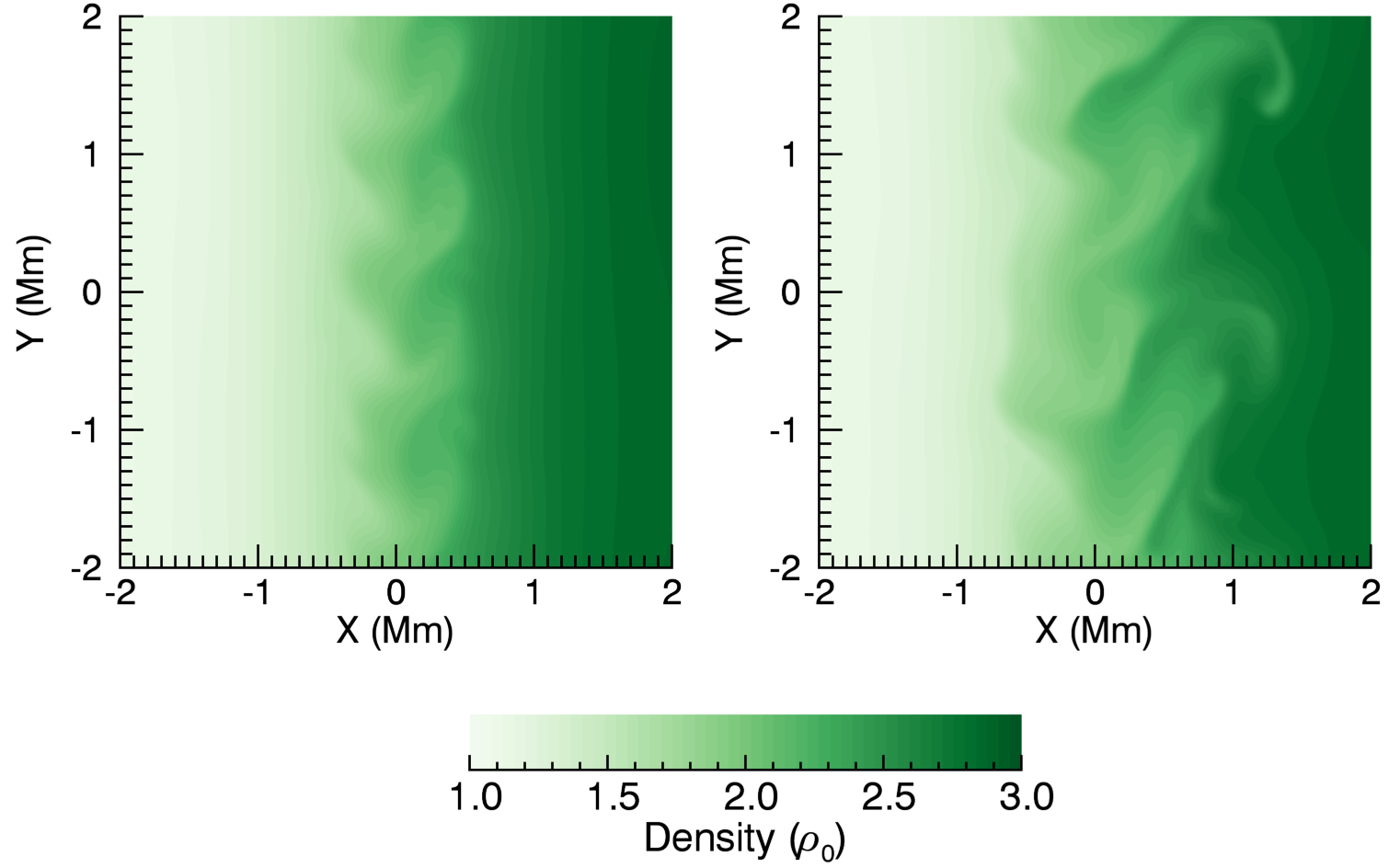}
  \caption{Density profile at $z=0$ Mm for $l = 20$ Mm (left) and $l=50$ Mm cases. We show the profiles at $t = 580$ s (left) and $t= 1450$ s (right), respectively. We note that due to the different field line lengths, these times both correspond to the same number of wave periods as panel (d) in Fig. \ref{khi_panels}.}
    \label{khi_length_panels}
\end{figure}

In Fig. \ref{khi_length_panels}, we show the disruption of the cross-sectional density profile for the $l=20$ Mm (left hand panel) and $l = 50$ Mm (right hand panel) cases. In terms of the number of driving periods (approximately $14.5 \tau$), both of these plots correspond to the same time as panel (d) in Fig. \ref{khi_panels}. The instability is well developed in both cases. As a result of the increased magnetic tension forces, we see that the Kelvin-Helmholtz mixing layer is much smaller for the shorter field line cases. As a result, observational signatures of the KHI are likely to be more difficult to detect in short coronal loops. Additionally, in non-ideal regimes, any significant wave heating will occur over a much narrower region for shorter field lines.

\subsection{Non-potential background field} \label{sec:npf}
Although the KHI is able to enhance the rate of energy dissipation, the total amount of wave energy available may be insufficient to balance expected coronal losses, particularly within active regions (for example, see Figs. \ref{cumulative_energy_flux} \& \ref{cumulative_energy_flux_line}). As we have seen, the existence of wave resonances will significantly enhance the amount of wave energy, however, it remains unclear whether these are sufficiently widespread and long-lived to meaningfully increase the energy injection rate. Despite this, even without sufficient wave energy, the dynamic instabilities may remain relevant in the context of the coronal heating problem as they may facilitate the release of energy stored in the background magnetic field. In particular, the compressive flows associated with the development of the KHI may trigger a faster rate of magnetic reconnection in the background field. In order to investigate this potential mechanism, in this section we adapt the model described above by including a shear in the background magnetic field. This provides an additional source of energy which may enhance inferred heating rates.

\begin{figure}[h]
  \centering
  \includegraphics[width=0.48\textwidth]{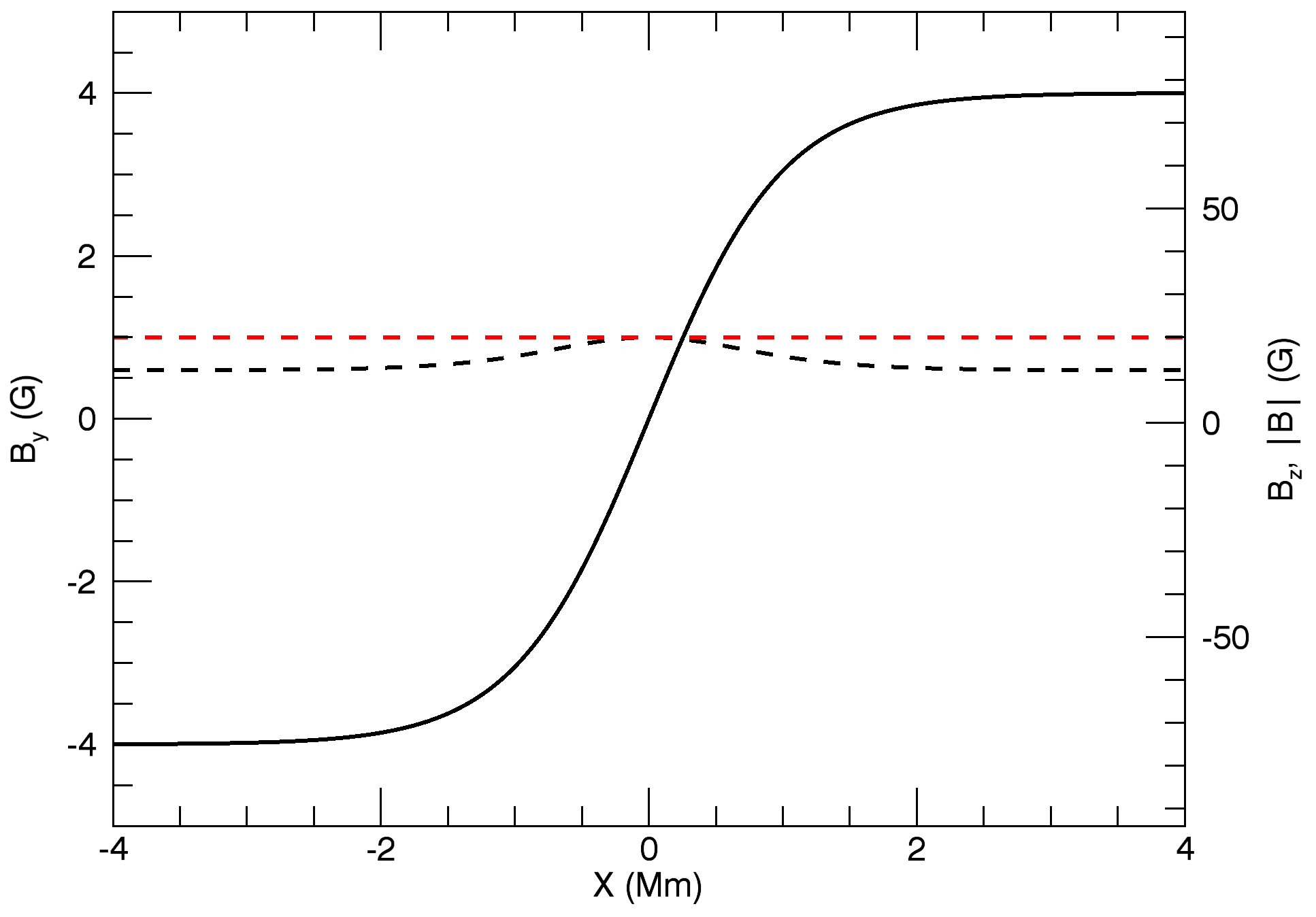}
  \caption{Transverse profiles of $B_y$ (solid black line), $B_z$ (dashed black line) and the total field strength (dashed red line) for the $\alpha = 0.2$ case. Here, the solid line uses the left hand axis and the dashed lines use the right hand axis.}
    \label{in_shear_field}
\end{figure}

In transversely oscillating coronal loops, the presence of magnetic twist will modify the formation and subsequent growth rate of the KHI \citep[e.g.][]{Howson2017A, Terradas2018} and even relatively weak twist may stabilise the system over the decay time of transverse loop oscillations \citep{Barbulescu2019}. For cylindrical coronal loops with a density enhancement, magnetic twist is modelled using a field component that is perpendicular to both the loop axis and the density gradient ($B_{\phi}$ in cylindrical geometry). In the Cartesian geometry considered here, this will correspond to a non-zero $B_y$ term in the equilibrium field. In the following, we consider $B_y$ defined as
\begin{equation} \label{mag_twist1}
B_y(x) = \alpha B_0 \tanh\left(\frac{x}{l_0}\right),
\end{equation}
where $-1 \le \alpha \le 1$ is a parameter controlling the strength of the magnetic shear and $B_0 = 20$ G is the magnetic field strength implemented in the uniform field simulations. We note that this profile reduces to the earlier setup for $\alpha = 0$. The constant, $l_0$, controls the width of the magnetic shear region. In order to maintain constant magnetic pressure and thus an initial equilibrium (there is no magnetic tension force in the background field), we define $B_z$ as
\begin{equation} \label{mag_twist2}
B_z(x) = \sqrt{B^2_0 - B^2_y(x)}.
\end{equation}
The inclusion of the shear component in the magnetic field introduces currents in the background field. These currents are field aligned and thus do not contribute any (initial) Lorentz force, however, they can contribute to plasma heating in non-ideal regimes. In order to investigate how this shear component in the magnetic field affects the energetics of the instability, we consider simulations with $\alpha \in \left\{0.05, 0.1, 0.2\right\}$. These values correspond to shear field strengths of 1, 2 and 4 G, respectively. The transverse profiles of $B_y$ (solid black line), $B_z$ (dashed black line) and the total field strength (dashed red line) for the $\alpha = 0.2$ case are shown in Fig. \ref{in_shear_field}. In this figure, the left hand axis corresponds to the solid line and the right hand axis corresponds to the dashed lines. In this section we have implemented $n_x = 256$ and $l = 100$ Mm. 

Since the magnetic field strength is unchanged throughout the domain, the local Alfv\'en speed profile remains the same as in the previous cases. However, the inclusion of a $B_y$ term effectively lengthens most of the field lines (not $x=0$, where $B_y = 0$) and thus modifies the natural Alfv\'en frequencies. This, in turn, affects the excited wave amplitudes (although not at $x=0$, in the linear regime). An additional consideration is that for the imposed wave driver, the velocity field is no longer perpendicular to the magnetic field. This results in a fraction of the injected energy exciting a slow wave \citep[see, for example, Fig. 10 in][]{Fyfe2020}. It is theoretically possible to impose a wave driver that satisfies $\vec{v} \cdot \vec{B} = 0$ everywhere on the lower $z$ boundary, however, this would require a precisely varying profile. This would inevitably differ across simulations and hence we do not consider this here.

At early times (prior to KHI formation), the cases with a shear component in the field evolve in a similar fashion to the potential field simulations. The phase mixing profile is modified slightly due to the change in the profile of the natural Alfv\'en frequency, however, the magnitude of the velocity shear across the resonant layer is largely unchanged. In the lower panel of Fig. \ref{something}, we compare the transverse profile of $v_y$ at $t=950$ s for the $\alpha = 0.2$ case (thin black line) with the $\alpha = 0$ simulations discussed above. The only significant difference is seen at large $|x|$, where the difference in field line frequencies between the simulations is greatest. We note that close to $x=0$ Mm, there is very little difference in the transverse velocity shear. 

With regards to the other components of the velocity field ($v_x$ and $v_z$), there are some differences between the simulations. The tilting of the field for $\alpha \ne 0$ cases inclines the natural oscillation mode of the Alfv\'en wave out of the horizontal plane and is thus associated with a velocity component in the $z$ direction. A component of $v_z$ also exists in the straight field case but this is a non-linear effect associated with the ponderomotive force \citep[e.g.][]{Hollweg1971, Tikhonchuk1995, Verwichte1999, Prok2019}. Additionally, the inclusion of background currents excites velocities in the $x$ direction, even in the linear regime. The perturbed field has a component of $B_y$ which interacts with $j_z$ in the background field to induce an $x$ component of the Lorentz force \citep[e.g.][]{Howson2019}. Again, the straight field cases can induce $x$ velocities (prior to the formation of the KHI) but this is a non-linear effect caused by phase mixing and the enhanced magnetic pressure on the resonant layer \citep[e.g.][]{Nakariakov1997, Botha2000, Thurgood2013}. Despite these differences, for the relatively weak magnetic twist here, wave dynamics are still dominated by the $v_y$ and $B_y$ components of the perturbed fields.

\begin{figure}[h]
  \centering
  \includegraphics[width=0.48\textwidth]{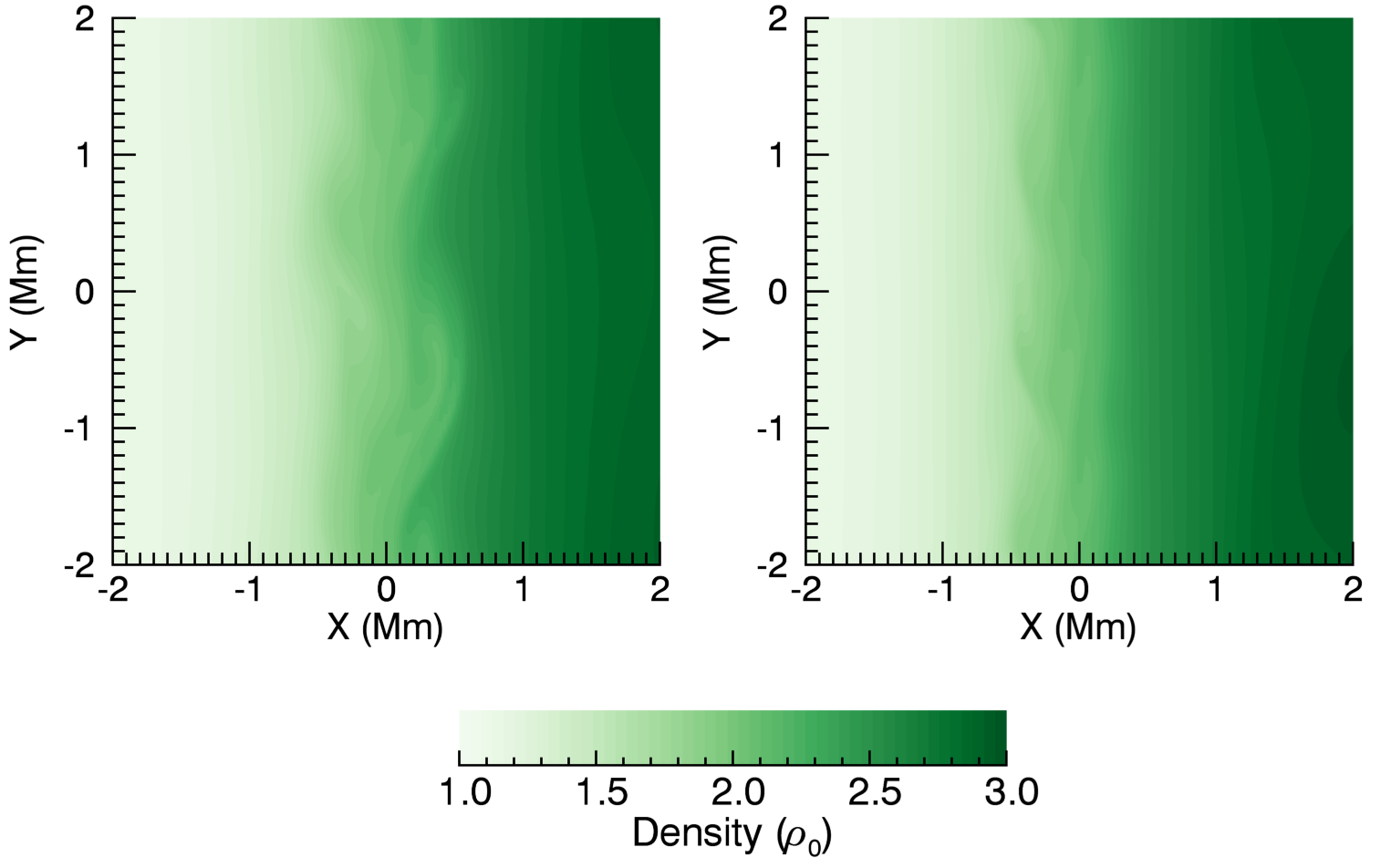}
  \caption{Density profile at $z=0$ Mm for thex $\alpha = 0.1$ (left) and $\alpha = 0.2$ (right) cases at $t \approx 2900$s.}
      \label{khi_field_shear_panels}
\end{figure}

Despite the minimal change in the velocity shear across the resonant layer for the non-potential cases, we see that the suppressive effect of magnetic twist (in cylindrical regimes) also applies here. In Fig. \ref{khi_field_shear_panels}, we show the density profile at $t \approx 2900$ s for the $\alpha = 0.1$ (left) and $\alpha = 0.2$ (right) cases. In terms of the number of wave periods ($\tau$), this is the same time as panel (d) in Fig. \ref{khi_panels} and both panels in Fig. \ref{khi_length_panels}. We see that increased field shear (right hand panel) reduces the size of the Kelvin-Helmholtz vortices and that both cases, show vortices with a much smaller spatial extent than the straight field case (compare with panel (d) in Fig. \ref{khi_panels}). Indeed in the $\alpha = 0.2$ case (where $|B_y|/|B|$ is still relatively small), there is very little deformation of the initial density profile.

\begin{figure}[h]
  \centering
  \includegraphics[width=0.48\textwidth]{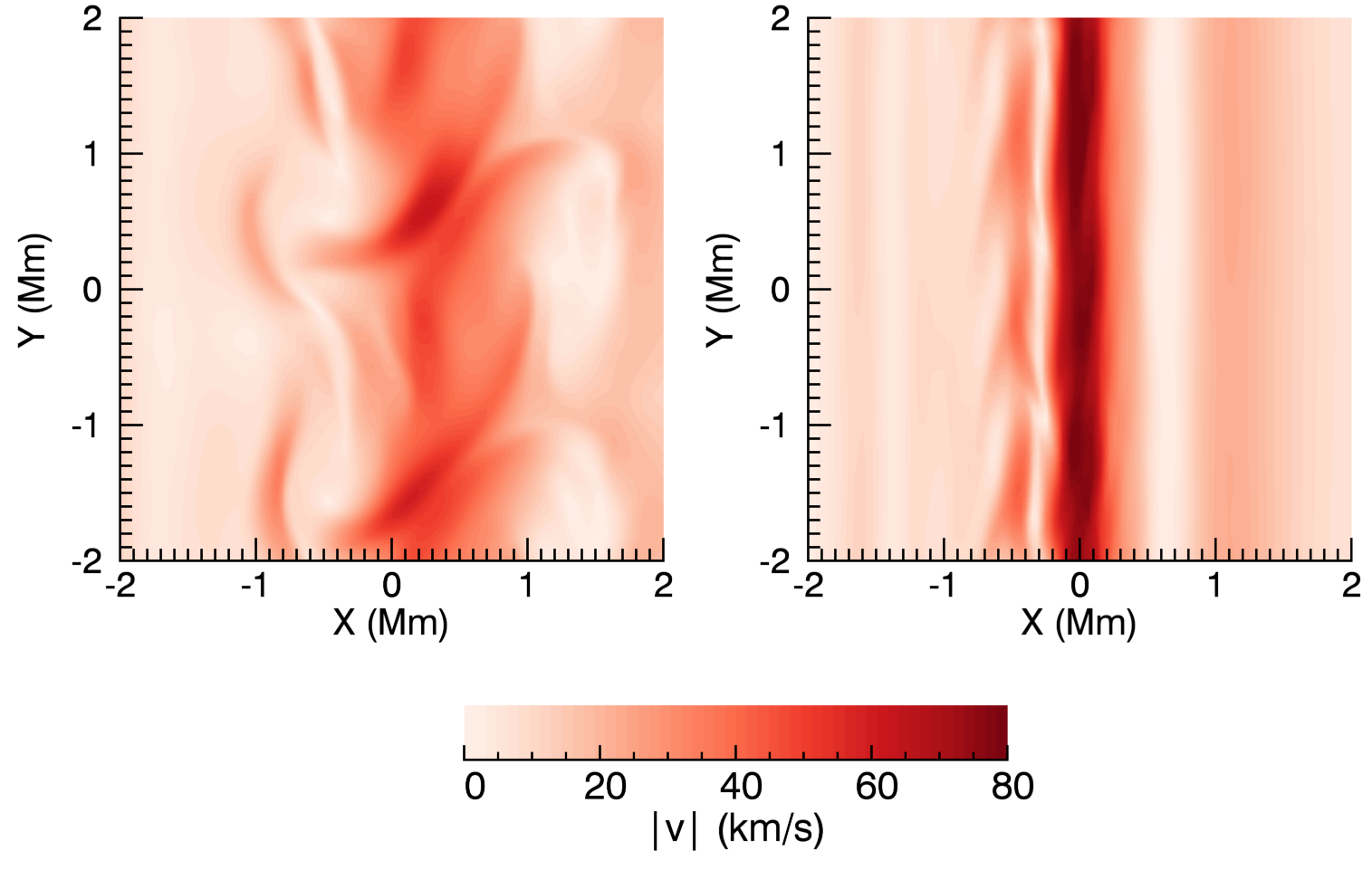}
  \caption{Profile of $|v_h| = \sqrt{v_x^2+v_y^2}$ for the straight field, $l=100$ Mm (left) and $\alpha = 0.2$ (right) cases at $t \approx 2900$s.}
    \label{khi_vel_panels}
\end{figure}

The slower instability growth rate and the reduced disruption in the higher $\alpha$ cases allow the resonant layer at $x=0$ to stay coherent for longer during the simulation runtime. This allows larger wave amplitudes to form on the resonant field lines in comparison to the straight field case, producing a greater transverse velocity shear. Despite this enhanced shear, the KHI growth rate remains suppressed (although not fully) for the non-potential field cases throughout the numerical experiments. In Fig. \ref{khi_vel_panels}, we compare the magnitude of the horizontal velocity field ($\sqrt{v_x^2+v_y^2}$) on the $z=0$ Mm plane for the $\alpha = 0$ (left hand panel) and $\alpha = 0.2$ (right hand panel) cases at $t \approx 2900$s. In the left hand panel, we see that the largest velocities form on the boundary of KHI vortices, driving the subsequent evolution of the instability. In the right hand panel, on the other hand, the Kelvin-Helmholtz vortices are much less apparent despite the presence of a greater velocity shear. It is apparent that the classical phase mixing pattern (corresponding to coherrent vertical banding in this topology), is disrupted for the straight field case but remains evident in the right hand panel.

In the third row of Fig. \ref{growth_panels}, we quantify the instability growth rates for the different field configurations and display the effects on the formation of small scales in the magnetic and velocity fields. In the left hand panel of the row, we show a proxy for the growth rate of the instability. In previous sections, we use the mean of $|v_x|$, however, this is no longer appropriate for the sheared field as an $x$ component in the velocity is directly excited by sheared field (as described above). Therefore, we use the maximum change in the plasma density throughout the numerical domain instead. This quantity is not perfect as the compressive nature of the wave driver and non-linear effects do modify the density profile even before the formation of the instability. However, as the initial Alfv\'en wave is largely incompressible and oscillates on planes of constant density, it still provides clear identification of the time when the first vortices begin to deform the density profile. In agreement with previous studies, we see that the inclusion of shear in the background magnetic field delays the onset of instability formation (by several wave periods in some cases) and reduces the subsequent growth rate. Once again, for clarity, we show a zoomed in version of the initial growth phase in the right hand panel of Fig. \ref{growth_panels_zoom}. In both versions of the plot, we see that the $\alpha = 0.2$ case (blue line), the effects of the KHI are difficult to detect in this quantity, with the most apparent signature being the breakdown in the oscillatory pattern at $t \approx 8 \tau$. 

As there are currents present in the background magnetic field (for $\alpha \ne 0$), to isolate the effects of the KHI, in the central panel of the third row in Fig. \ref{growth_panels}, we display the change in the mean (solid lines) and the change in the maximum (dashed lines) currents. We see that increasing shear reduces the magnitude of the Kelvin-Helmholtz currents that form. This is a little misleading as the sheared cases have significantly larger background currents. Indeed, for the $\alpha = 0.2$ case, the increase in the mean current only represents a 20\% enhancement on top of the background. However, in these simulations, any Ohmic dissipation associated with the KHI would be small in comparison to that associated with the diffusion of the background field note that this assumes a spatially and temporally uniform value of resistivity, while in reality numerical dissipation will be enhanced in the current layers associated with the KHI. Despite the inclusion of background currents in the $\alpha \ne 0$ cases, the largest currents (although highly localised) are observed in the $\alpha = 0$ simulation, as a result of the most disruptive vortices.

In the lower right hand panel of Fig. \ref{growth_panels}, we show the evolution of the mean and maximum vorticities. Unlike with the currents, the largest values of both the means and the maxima occur in the more sheared cases. Prior to the evolution of the instability, the vorticity evolves in a very similar manner, as phase mixing is progressing in the same fashion across simulations. However, this changes as the resonant layer of field lines is disrupted as the KHI forms. The delayed onset of the instability in the higher $\alpha$ cases allows for a greater increase in the wave amplitude on the resonant $x=0$ plane and a prolonged period of classical phase mixing \citep{Heyvaerts1983}. This permits enhanced vorticities in the sheared cases in comparison to the $\alpha = 0$ simulation. As with the previous simulations, the mean vorticity decreases during the growth of the instability. Conversely, we see that the maximum initially increases, but note that this is highly localised around the edges of the largest Kelvin-Helmholtz vortices.

\subsection{Magnetic reconnection}
Due to the high magnetic Reynolds number expected in the corona, magnetic field lines are approximately frozen into the plasma. As such, as the KHI develops, field lines will be carried with the vortices that are apparent in the density profile and the velocity field. In particular, the disruptive flows associated with the instability can create large misalignments between neighbouring magnetic flux bundles, as shown by the sharp growth in the peak current density. In the presence of a finite resistivity this may trigger magnetic reconnection. For potential fields, this process can only release magnetic energy injected by the wave driver. However, for non-potential fields it may trigger or enhance the rate of reconnection of the background magnetic field. In this section, we evaluate the rate of magnetic reconnection in each of the simulations described above.

\begin{figure}[h]
  \centering
  \includegraphics[width=0.48\textwidth]{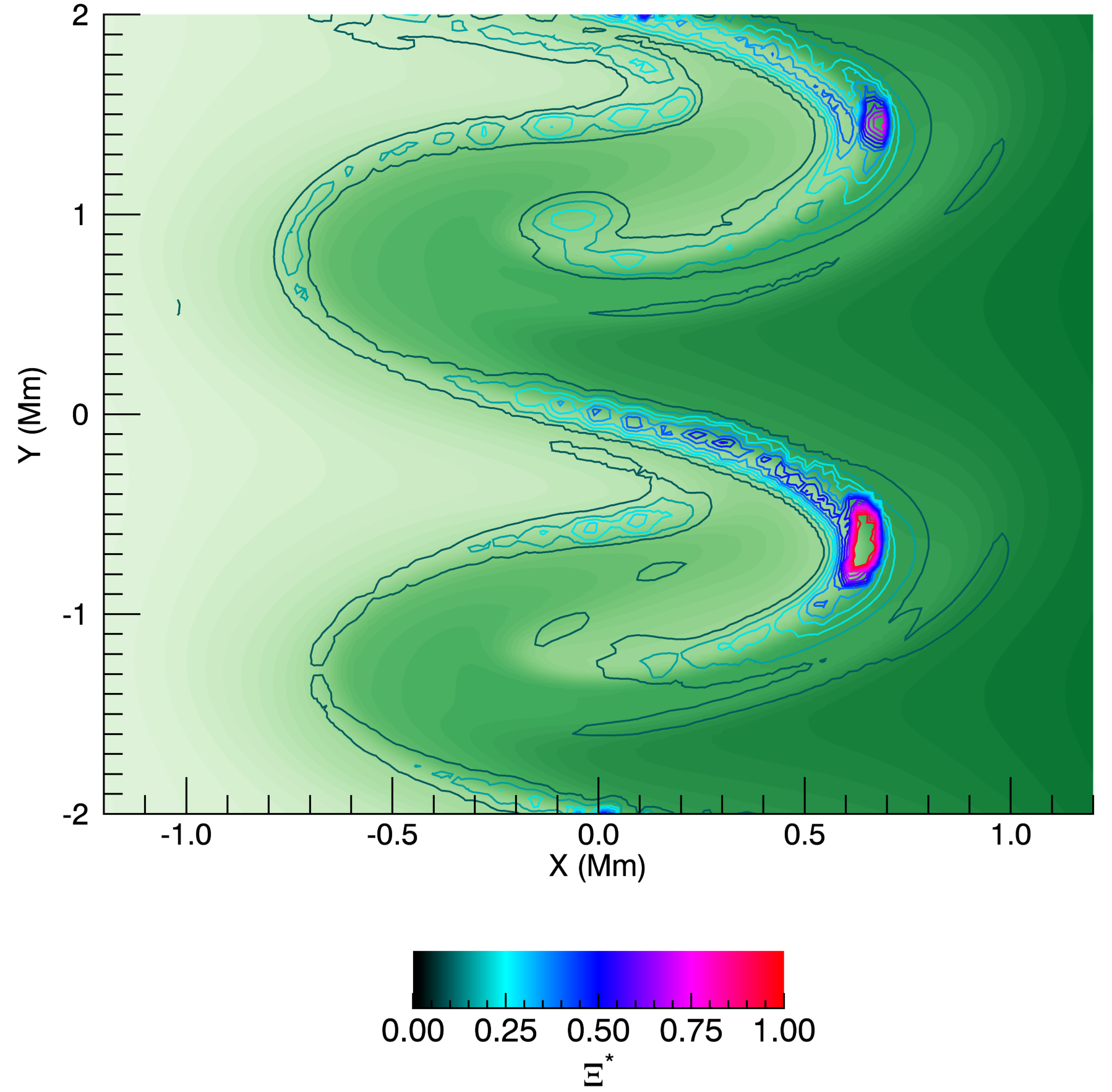}
  \caption{{\emph{Filled green contours:}} Density profile at $z = 0$ Mm. The colour table used for the density is the same as in Figs. \ref{khi_panels}, \ref{khi_length_panels} and \ref{khi_field_shear_panels}. {\emph{Line contours:}} Value of $\Xi^*$ (see eq. \ref{eq:mag_rec_numerical}) for field lines embedded within the Kelvin-Helmholtz mixing layer. The time shown is $t \approx 1090$ s for the straight field, high resolution, $l = 100$ Mm simulation. We have normalised using the maximum value of $\Xi^*$ at this time. For clarity, we have restricted the $x$ axis in each panel to $-1.2 \text{ Mm} \le x \le 1.2$ Mm.}
\label{dens_rec_contour}
\end{figure}

Following the injection of energy by the driver (but before instability onset) the plane $x=0$ is both a magnetic shear layer and velocity shear layer. The interplay of different instabilities and associated energy release at such structures has been examined in detail is studies by \citet{Einaudi1986,Dahlburg1997}. Indeed, \citet{Dahlburg1997} argue that the magnetic shear fundamentally alters the nature of the KHI by allowing magnetic reconnection to become important.
As discussed by \citet{Wyper2013}, quantifying and interpreting reconnection in such a configuration is highly challenging, as the KHI induces multiple localised reconnection processes in current layers within the domain. In three-dimensional non-ideal MHD, a necessary and sufficient condition for magnetic reconnection to occur is
\begin{equation} \label{eq:mag_rec}
\Xi = \int_l E_{\parallel} \,\,\mathrm{d}s \ne 0,
\end{equation}
where we have $E_{\parallel} = \eta j_{\parallel}$ and the integral is calculated along magnetic field lines \citep[e.g.][]{Schindler1988}. Within a particular non-ideal region, the maximum value of the integral in eq. \ref{eq:mag_rec} will give the rate of magnetic reconnection. When multiple, distinct regions of non-zero $\Xi$ exist, a more nuanced approach to calculating the global reconnection rate is required. Equation 21 in \citet{Wyper2015} gives the global reconnection rate in the presence of fragmented current regions as 
\begin{equation}
\lvert \Xi_{\max}\rvert+ \sum \lvert \Xi_{\text{l.m.}} - \Xi_{\text{s.p.}} \rvert.
\end{equation}
This requires the identification of the integral at local maxima, $\Xi_{\text{l.m.}}$, and at distinct saddle points between maxima $\Xi_{\text{s.p.}}$ \citep[see][for more details]{Wyper2015}. The summation is calculated over local maxima. This quantifies the net rate of change of flux connectivity with respect to the foot points at either ends of our domain ($z=\pm z_{max}$). For the simulations described above, since we have $\eta = 0$, we can consider $E_{\parallel} = 0$. However, due to the finite size of numerical grid cells, the effective $\eta$ will be non-zero and thus we can qualitatively investigate the numerical reconnection rate using 
\begin{equation} \label{eq:mag_rec_numerical}
\Xi^* = \int_l j_{\parallel} \,\,\mathrm{d}s.
\end{equation}
Although the exact value of the numerical resistivity is not known, we can compare approximate reconnection rates between simulations (with the same numerical resolution) under the assumption that the true reconnection rate is proportional to 
\begin{equation} \label{eq:rec_rate_numerical}
\lvert \Xi^*_{\max}\rvert+ \sum \lvert \Xi^*_{\text{l.m.}} - \Xi^*_{\text{s.p.}} \rvert.
\end{equation}
A more thorough investigation is described in Sect. \ref{non_id_reg} using an explicitly defined $\eta$, instead.

In Fig. \ref{dens_rec_contour}, we begin by showing the value of $\Xi^*$ (unfilled contours) overplotted on the density profile (filled green contours) at $t \approx 1090$ s for the $\alpha = 0$, $n_x=256$, $l = 100$ Mm simulation. In order to calculate the integral, $10^4$ magnetic field lines were traced from a grid on the lower $z$ boundary with start points, $(x_s, y_s)$ satisfying -1 Mm $<x_s<1$ Mm and -2 Mm $< y_s <$ 2 Mm. We note that the wave driver displaces the magnetic field and thus the same field lines are not traced throughout the simulation. Instead, we simply select a sample of field lines which are ultimately embedded in the Kelvin-Helmholtz vortices.

In order to create Fig. \ref{dens_rec_contour}, the location of each field line was defined as its coordinate at $z=0$ (simulation midplane). We see that the largest values of the integral coincide with the boundaries of Kelvin-Helmholtz vortices where the velocity field is shearing the magnetic field lines. Meanwhile, the value of the integral is much smaller in the rest of the cross-section. In the locations of large $\Xi^*$, the velocity gradient, and hence the field shearing, is maximal. As such, the global reconnection rate will be dominated by small regions within the mixing layer.

\begin{figure}[h]
  \centering
  \includegraphics[width=0.48\textwidth]{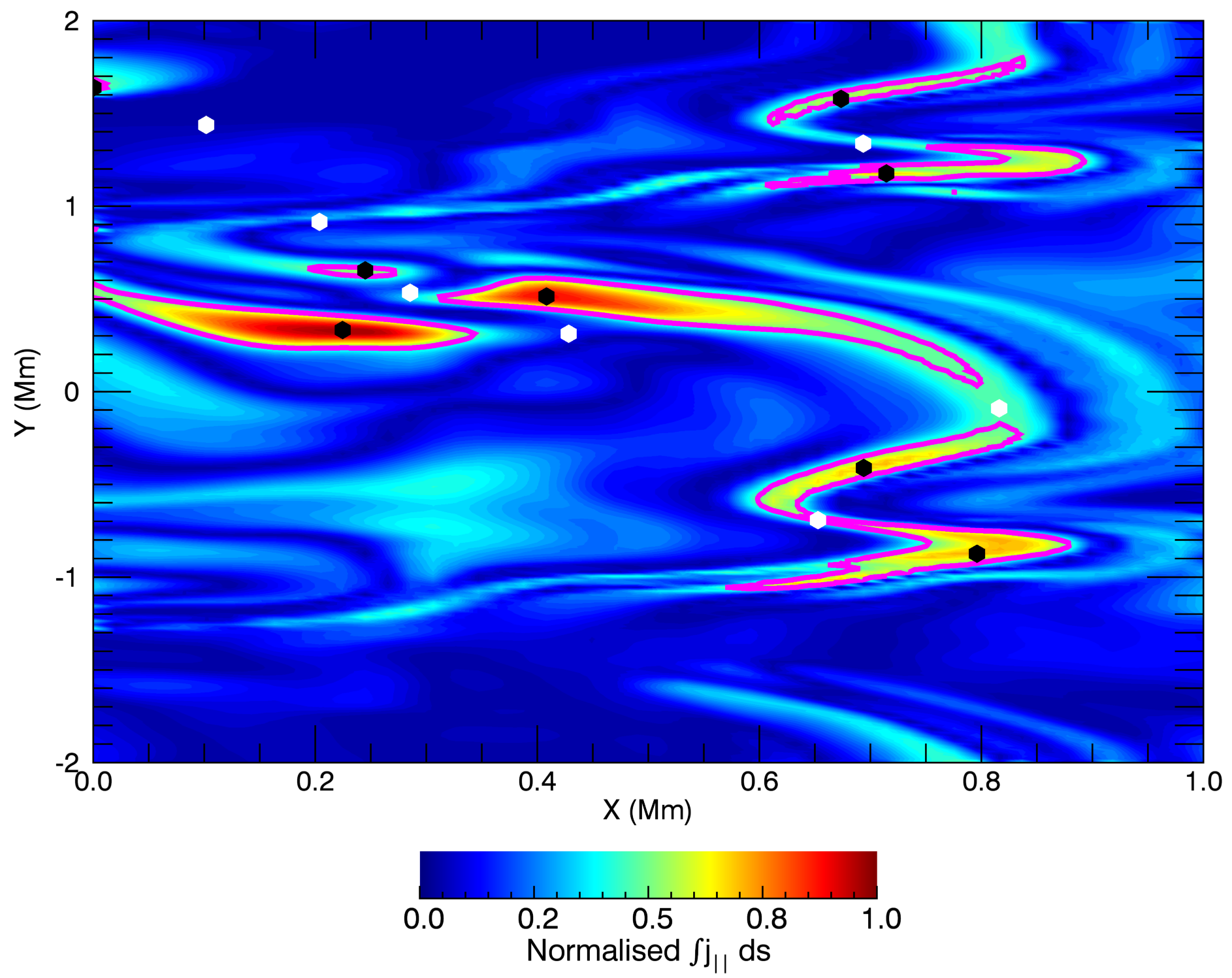}
  \caption{Contours show the parallel component of the current integrated over field lines traced from the lower $z$ boundary. Any given location corresponds to the position of the foot point of a magnetic field line. The pink line shows a level equal to half of the maximum value of the integral (0.5). Local maxima within the pink contour are identified with black points and the loci of associated saddle points are indicated with white points.}
    \label{rec_contour}
\end{figure}

Following the onset of the KHI, the spatial profile of $\Xi^*$ can be very noisy and thus many local maxima exist. In order to use eq.\ref{eq:rec_rate_numerical}, we must then also locate each of the saddle points linking these maxima. In Fig. \ref{rec_contour}, we depict the process undertaken for estimating the instantaneous reconnection rate. The contours show the value of $\Xi^*$ at $t= 1750 $ s in the case $n_x = 256$, $l=100$ Mm, $\alpha = 0$. Due to the large number of local maxima, in order to illustrate the method, we only show the peaks above a certain threshold (pink line). This is given by half of the maximum value of the integral at this time. However, in practice all of the local maxima and associated saddle points are identified. In Fig. \ref{rec_contour}, the eight black dots show the locations of local maxima within the distinct regions contained within the pink contour. The seven white dots show the associated saddle points. Thus to estimate the reconnection rate at this time, we use the values of $\Xi^*$ at these points (plus all other maxima and saddle points omitted from the figure) in eq.\ref{eq:rec_rate_numerical}.

\begin{figure*}[h]
  \centering
  \includegraphics[width=0.9\textwidth]{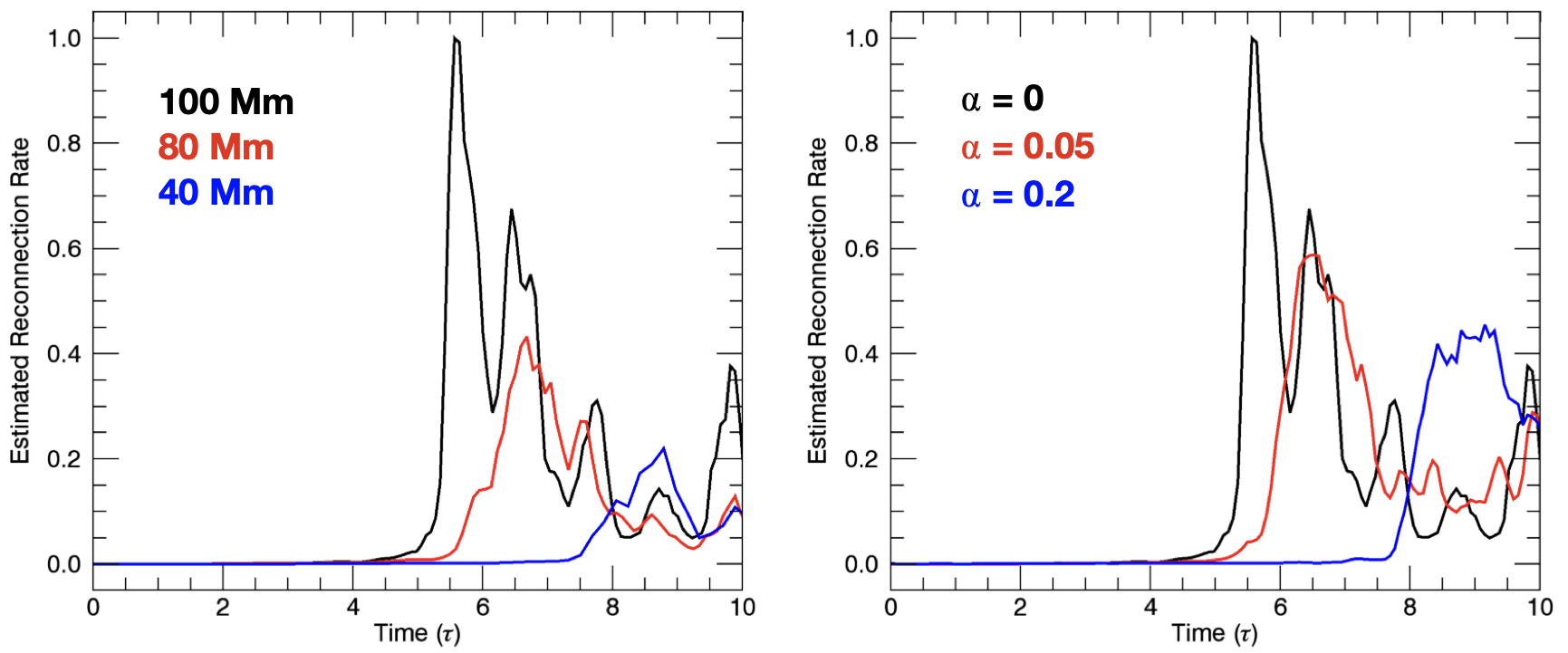}
  \caption{Estimated reconnection rate calculated using eq. \ref{eq:rec_rate_numerical}. In the left hand panel, we show the effects of field line length using simulations with $l=100$ Mm (black), $l=80$ Mm (red) and $l=40$ Mm (blue). In the right hand panel, we show the effects of shear in the background field using the simulations with $\alpha=0$ (no shear; black), $\alpha=0.05$ (low shear; red) and $\alpha=0.2$ (high shear; blue). Here we have normalised the rates by the maximum of the 100 Mm, $\alpha = 0$ case (black line).}
    \label{j_par_growth}
\end{figure*}

In Fig. \ref{j_par_growth}, we show how this estimated reconnection rate changes as a function of time (wave periods, $\tau$) for different field line lengths (left hand panel) and for different values of shear in the background field (right hand panel). For clarity, we have not included all simulation results and we have normalised all curves by the maximum value in the $l= 100, \, \alpha = 0$ case. We note that the two black curves correspond to the same simulation. In all cases, we see that the reconnection rate rises dramatically at the onset of the instability. As such, in terms of the number of wave periods, this increase is delayed when the KHI forms later (for shorter field lines and for $\alpha \ne 0)$. 

\begin{figure*}[h]
  \centering
  \includegraphics[width=\textwidth]{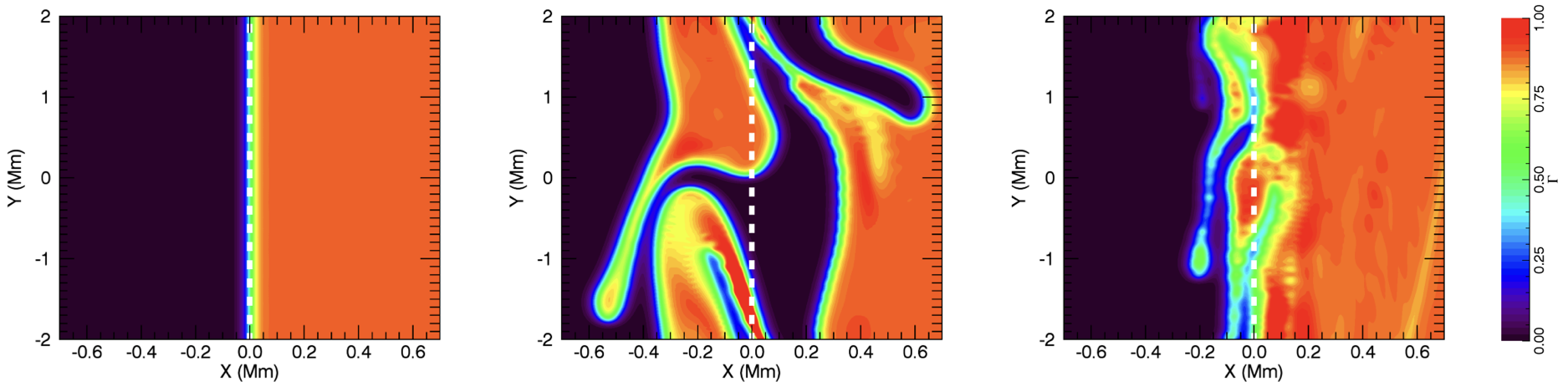}
  \caption{Density of traced field line end points calculated using $\Gamma$ (see eq. \ref{Gamma}) and normalised to the maximum value. This shows reconnection of field lines across the resonant plane. Field lines were traced from $x \ge 0$ on the lower $z$ boundary. We show the $\alpha = 0$ case at $t=0$ s (left hand panel) and at $t= 1090$ s (centre panel) and the $\alpha = 0.2$ case at $t=1740$ s (right hand panel). The white dashed line corresponds to the initial position of the resonant plane.}
    \label{heat_map}
\end{figure*}

In addition, we see that when the size of the Kelvin-Helmholtz vortices is reduced, we calculate a lower reconnection rate. For example, the peaks of the blue curves (left; $l=40$ Mm, right $\alpha = 0.2$) are much lower than the unsuppressed $\alpha = 0$ case (black curve). We note that in terms of the reconnection rate, the free magnetic energy for the $\alpha \ne 0$ cases is not able to fully compensate for the reduced instability growth rate. However, even though the density deformation is very small in the $\alpha = 0.2$ case (see right hand panel of Fig. \ref{khi_field_shear_panels}), we do see that the KHI significantly enhances the reconnection rate above the level derived from the equilibrium field. As such, the KHI may be energetically important even if it is difficult to observe in the corona.

The reconnection can be tracked by following the connectivity of field lines embedded within the Kelvin-Helmholtz vortices. In Fig. \ref{heat_map}, we show how the connectivity changes. To create this plot, we traced $1.6 \times 10^5$ field lines from a set of foot points on the lower boundary. In this case, we restricted the start points to 0 Mm $< x_s <$ 1 Mm and -2 Mm $< y_s <$ 2 Mm. In particular, the $x$ coordinates of the start points were all on one side of the resonant layer. Initially, the $x$ component of the magnetic field is 0 and therefore (at the start of the simulations), the field lines are each confined to a particular value of $x$ (although the value of $y$ does change in the $\alpha \ne 0$ cases). The KHI causes reconnection across the shear layer and thus, some field lines are ultimately connected to points with $x$ coordinates that are less than 0. To show this, we display a measure of the density of traced field line end points (see eq. \ref{Gamma} below). This is caluclated by first computing all pairwise distances, $d_{x,y}^p$, between points $(x,y)$ on the upper boundary and the end points, $p$. For example, at every point on the upper boundary, we calculate all the distances between this point and the traced field line end points. We then calculate a variable, $\Gamma = \Gamma(x,y)$, as
\begin{equation} \label{Gamma}
\Gamma = \sum_{p} \exp\left\{-\left(\frac{d^p_{x,y}}{\lambda}\right)^2\right\},
\end{equation}
where the summation is over the traced field lines and $\lambda$ is a length scale set to be ten times the separation between neighbouring field line start points (in order to generate a smooth measure). We display $\Gamma$ at $t=0$ (left hand panel) and $t=1090$ s (central panel) for the $\alpha = 0$ case, and at $t= 1740$ s for the $\alpha = 0.2$ case (right hand panel). The central and right hand panels correspond to times when the instability is developing in their respective simulations. In each case, we have normalised by the maximum value of $\Gamma$ as the relative size is unimportant. The dashed white line shows the initial location of the resonant plane.
 
In the left hand panel, we see that initially the field lines do not cross the resonant plane ($B_x = 0$ everywhere). However, at later times, once the instability has developed, field lines have reconnected across the mixing layer, in both the $\alpha = 0$ and $\alpha = 0.2$ cases. This leads to an increase in field line density (of traced field lines) for $x < 0$ Mm and a corresponding decrease in field line density for $x > 0$ Mm. As the wave driver only moves foot points in the $y$ direction, and the velocity on the upper boundary is forced to 0, any field lines that connect to $x<0$ at the upper boundary must have changed their connectivity. By comparing the central and right hand panels, we can clearly see the reduced size of the mixing layer. Field lines reconnect over a much smaller distance for the sheared field case (right hand panel).
 
\subsection{Non-ideal regimes} \label{non_id_reg}

Investigating the Kelvin-Helmholtz instability in an explicitly non-ideal regime is difficult because of the suppressive effects that dissipation has on the growth rate of the instability \citep[e.g.][]{Howson2017B}. In this context, it is important to note that we cannot accurately replicate the high Reynolds number coronal conditions in 3D MHD simulations. As such, even the ideal simulations discussed above will exhibit artificially low growth rates. On account of these considerations, in this section we investigate the effects of a local resistivity term (non-uniform in space and time) that is triggered by high currents. By choosing a suitable threshold for this critical current, we allow the instability to form as above, before the resistivity is activated.

We consider the $\alpha = 0$ and $\alpha = 0.1$ cases, as these allow the same threshold current to be selected without significantly impeding the initial growth of the instability in either case. We selected a threshold of $5 \times 10^{-4} \text{ A m}^{-2}$ and a critical resistivity corresponding to a magnetic Reynolds number of approximately $10^3$.  

\begin{figure*}[h]
  \centering
  \includegraphics[width=0.97\textwidth]{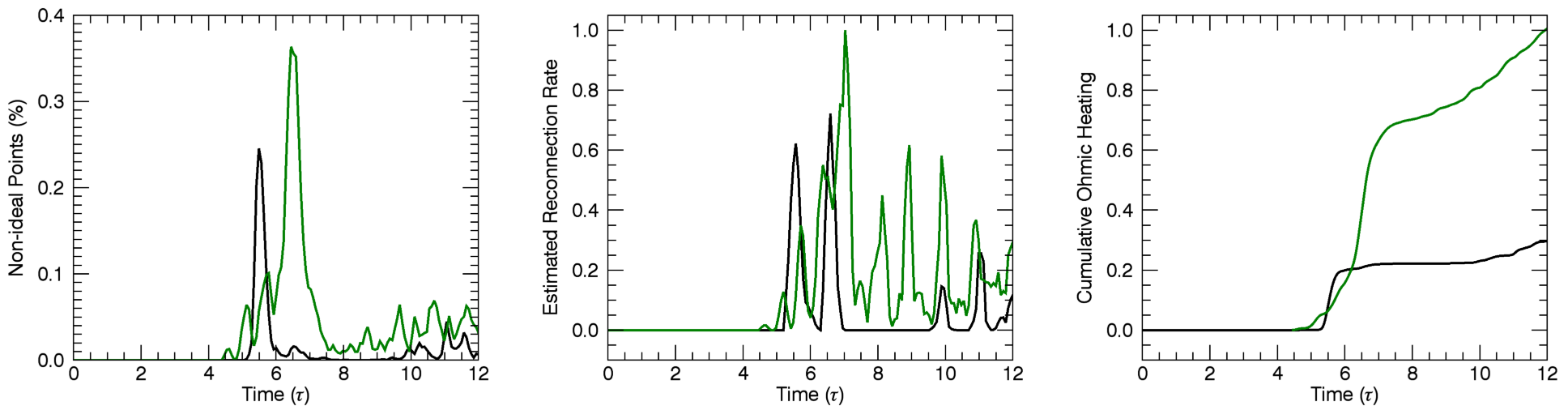}
  \caption{{\emph{Left:}} Percentage of grid cells at which the critical resitivity is triggered. {\emph{Centre:}} Global magnetic reconnection rate as a function of wave period. {\emph{Right:}} Volume integrated Ohmic heating rate. We show the $\alpha = 0$ (black) and $\alpha = 0.1$ (green) cases. For the central and right hand panels, we have normalised by the maximum of the blue curve.}
    \label{anom_resist_plots}
\end{figure*}

In Fig. \ref{anom_resist_plots}, we show a summary of the results of these two simulations. For all plots, the black curve shows the $\alpha = 0$ case and the green curve shows the $\alpha = 0.1$ case. In the left hand panel, we  show the percentage of grid points in the domain where the critical current is triggered. We note that for both simulations, the number of points remains small ($< 1\%$) at all times and peaks during the steep growth phase of the instability. The threshold is exceeded at a small number of points just before the instability onset time in the $\alpha = 0.1$ case. The presence of the background currents (together with phase mixing currents) in the $\alpha = 0.1$ case means that the critical current threshold is surpassed before the onset of the KHI in the sheared field simulation. This is not the case for the $\alpha = 0$ simulation. 

In the central panel of Fig. \ref{anom_resist_plots}, we show the reconnection rate for the two simulations as a function of wave periods. This is calculated as described above, except we now use $E_{\parallel}$ instead of $j_{\parallel}$. In other words, unless a field line passes through a grid cell where the critical resistivity is activated, the integral will be zero. For this figure, we have normalised both curves by the maximum of the green curve. As before, we see a sharp rise in the reconnection rate in both cases as the instability develops. However, in contrast to the simulations described above, we see that the sheared field simulation typically has a higher reconnection rate than the $\alpha = 0$ case. This is simply due to the increased number of grid points at which the critical resistivity is triggered. The highly variable nature of these curves is caused by the localised and intermittent activation of this resistivity due to formation and decay of individual current sheets during the development of the KHI.

Similarly, the volume integrated Ohmic heating rate (right hand panel of Fig. \ref{anom_resist_plots}), is larger for the non-potential field case. Again, this can be attributed to the greater number of grid cells at which the critical resistivity is activated. These results show that it is possible to select a resistivity profile such that energy release during the instability will be greater in sheared field cases even though the KHI growth rate is lower. This result will be physically relevant if the small scales that develop turbulent-like flows that develop following the onset of the instability trigger a regime with a higher rate of resistive dissipation.
 
\section{Discussion and conclusions} \label{Discussion}
In this article, we have presented the results of MHD simulations investigating the formation of the magnetic Kelvin-Helmholtz instability. Using a fixed frequency sinusoidal velocity driver, we introduced shear Alfv\'en waves into a coronal model containing a cross field gradient in the local Alfv\'en speed. Following a period of phase mixing and the growth of a resonant standing wave, the transverse velocity shear became unstable. As a result, small scales form in both the velocity and magnetic fields, which will lead to enhanced viscous and Ohmic heating rates in non-ideal regimes. Even for high (but finite) Reynolds numbers, the field-aligned currents that form as a result of the instability will be associated with magnetic reconnection. We see that the instability triggers reconnection in all cases and the calculated rate is highest when the growth rate is largest. 

In terms of magnetic reconnection, for the current setup, the extra energy associated with a non-potential shear component in the background field is not sufficient to compensate for the low growth rate. For the simulations described here, both the shear in the background field (for $\alpha \ne 0$) and the out-of-phase waves (before the onset of the KHI) are associated with a gradient ($\partial{B_y}/\partial{x}$) across the resonant field lines. Coincidentally, at the instability onset time, the amplitude of the perturbed field (on the resonant plane) is comparable to the size of the shear component in the $\alpha = 0.2$ case. Furthermore, the width of the background current layer (see Fig. \ref{in_shear_field}) is similar to the width of the phase mixing layer (see lower panel of Fig. \ref{something}). As such, as the instability threshold is reached, the largest currents in the non-potential case are approximately twice those in the potential field simulations (wave currents and background currents in the $\alpha = 0.2$ case and just the wave currents in the $\alpha = 0$ case). The disruptive KHI flows dynamically generate field line misalignments over small length scales, and thus further enhance the current density. As the potential field case develops much larger Kelvin-Helmholtz vortices, field lines with a much greater initial separation are forced together (more so than in the $\alpha \ne 0$ cases). As a result, even though the pre-instability gradients are smaller for the $\alpha = 0$ simulation, the greater disruption that we see in this case is more than sufficient to compensate. As such, this ultimately produces a higher reconnection rate when the field is not sheared.

In the sheared field cases, the background field varies over a length scale comparable to the initial density profile. However, it would be possible to use a magnetic field which varies over much smaller scales. An extreme example of this would be a step-change in the shear component of field across the resonant layer (an infinite current). In such a regime, the Kelvin-Helmholtz vortices may struggle to form due to the shear and the (zero) numerical resolution across the shear layer. However, even a very small disturbance in this case may well trigger significant reconnection (e.g. alongside the tearing instability). An analytic treatment of such a model may yield interesting in a future study. 

We also note that the observed reconnection rate will scale with imposed transport coefficients, e.g. magnetic diffusivity. In more diffusive simulations the growth rate is reduced and the KHI may be suppressed entirely \citep[see][for an investigation into this effect in coronal loops]{Howson2017B}. In this case, plasmas with higher Reynolds numbers would allow the instability to form sooner. This would mean that the amplitude of the resonant wave and, thus, the pre-instability currents may be different. With this in mind, the ratio between the background currents (from the magnetic shear) and wave currents need not be constant for different values of the resistivity. As such, the relationship between the reconnection rates obtained in the unsheared and sheared simulations may change too.

The true complexity of the coronal field remains unclear. As such, we cannot say how much magnetic shear is required to accurately represent coronal conditions. Indeed, it may be the case that the results here are not directly applicable to other field configurations. In particular, any field that does not suppress the growth rate of the instability need not exhibit the lower reconnection rates seen in the sheared cases. It is important to note that it is not specifically the free energy in the field that reduces the growth rate, but rather the geometry of the field employed here.

Despite this, if the corona is threaded by complex and intricate current sheets, our $\alpha = 0.2$ case may be the most appropriate of the simulations presented here. Although the instability growth rate is reduced and the density profile deformation is much less profound (and thus the instability would be difficult to detect directly) in this simulation, the KHI does still enhance the rate of reconnection beyond that apparent in the background field (see Fig. \ref{j_par_growth}). Therefore, it is possible that the KHI is relevant as a trigger for local reconnection events which could potentially lead to more widespread energy release \citep[e.g. due to an MHD avalanche][]{Hood2016, Reid2018}.

{\emph{Acknowledgements.}} The authors would like to thank the anonymous referee for providing comments and suggestions that significantly improved this article. The authors would also like to thank Mr Daniel Chambers forf his field line tracing codes. The research leading to these results has received funding from the UK Science and Technology Facilities Council (consolidated grant ST/N000609/1), the European Union Horizon 2020 research and innovation programme (grant agreement No. 647214). IDM received funding from the Research Council of Norway through its Centres of Excellence scheme, project No. 262622. This work used the DiRAC Data Analytic system at the University of Cambridge, operated by the University of Cambridge High Performance Computing Service on behalf of the STFC DiRAC HPC Facility (www.dirac.ac.uk). This equipment was funded by BIS National E-infrastructure capital grant (ST/K001590/1), STFC capital grants ST/H008861/1 and ST/H00887X/1, and STFC DiRAC Operations grant ST/K00333X/1. DiRAC is part of the National e-Infrastructure.

\vspace{1cm}

\noindent Reproduced with permission from Astronomy \& Astrophysics, \textcopyright \, ESO

\bibliographystyle{aa}        
\bibliography{KHI.bib}           

\end{document}